\UseRawInputEncoding
\documentclass[american,aps,pra, reprint, superscriptaddress]{revtex4-2}
\usepackage{amsthm}
\usepackage{amsmath}
\usepackage{array}

\usepackage{bm}
\usepackage{amssymb}
\usepackage{graphics,graphicx}
\usepackage{tikz}
\usepackage{float}
\usepackage{diagbox}
\usepackage{rotating}
\usepackage[unicode=true,pdfusetitle, bookmarks=true,bookmarksnumbered=false,bookmarksopen=false,
breaklinks=false,pdfborder={0 0 0},backref=false,colorlinks=false]
{hyperref}
\hypersetup{colorlinks,linkcolor=myurlcolor,citecolor=myurlcolor,urlcolor=myurlcolor}

\makeatletter
\usepackage{times}
\usepackage{braket}
\usepackage{colortbl}

\usepackage{array}
\newcolumntype{P}[1]{>{\centering\arraybackslash}p{#1}}

\definecolor{myurlcolor}{rgb}{0,0,0.7}

\makeatother

\newcommand{\ketbra}[2]{|{#1} \rangle \langle {#2} |}
\usepackage{times}

\newcommand{\mi}{\mathrm{i}}

\newtheorem{lemma}{Lemma}

\begin{document}
%\title{Classical communications with $N$ completely depolarizing channels with SWITCH}
\title{Classical communication with indefinite causal orders for $N$ completely depolarizing channels}

\author{Sk Sazim}
\email{sk.sazimsq49@gmail.com}
\affiliation{RCQI, Institute of Physics, Slovak Academy of Sciences, 845 11 Bratislava, Slovakia}
%\affiliation{QIC Group, Harish-Chandra Research Institute, HBNI, Allahabad, 211019, India}

\author{Michal Sedlak}
\email{michal.sedlak@savba.sk}
\affiliation{RCQI, Institute of Physics, Slovak Academy of Sciences, 845 11 Bratislava, Slovakia}

\author{Kratveer Singh}
%\email{kratveer050@gmail.com}
\affiliation{Indian Institute of Science Education and Research, Bypass Road, Bhauri, Bhopal 462066 India}

\author{Arun Kumar Pati}
\affiliation{QIC Group, Harish-Chandra Research Institute, HBNI, Allahabad, 211019, India}

\date{\today}

\begin{abstract}
%In the presence of indefinite causal order, 
If two identical copies of a completely depolarizing channel are put into a superposition of their possible causal orders, they can transmit non-zero classical information. Here, we study how well we can transmit classical information with $N$ depolarizing channels put in superposition of $M$ causal orders via quantum SWITCH. We calculate Holevo quantity if the superposition uses only cyclic permutations of channels and find that it increases with $M$ and it is independent of $N$. For a qubit it never reaches $1$ if we are increasing $M$. On the other hand, the classical capacity decreases with the dimension $d$ of the message system. 
Further, for $N=3$ and $N=4$ we studied superposition of all causal orders and %separately also 
uniformly superposed 
%various uniform superpositions of 
causal orders 
%involving terms 
%from 
belonging to different cosets created by cyclic permutation subgroup. 
\end{abstract}

%In the presence of indefinite causal order, two identical copies of a completely depolarizing channel (CDC) can transmit non-zero information. This effect emerges due to the quantum superposition of two alternative orders of these channels. Here, we study how well we can transmit classical information with superposition of $N$ depolarizing channels in a number of causal orders and find that 
%there can always be a nonzero classical communication rate. 
%% --- Reformulate a bit - Miso will do%%%%
%\tb {We uncover a simple yet crucial structure in the causal orders when we consider $N$ ($\geq 3$) CDCs in a quantum SWITCH. We find that there exist $N!$ causal orders for $N$ CDCs, and we can group them in $(N-1)!$ cosets of cyclic causal orders. And we discover that a single coset performs almost equally to all of them combined. However, a substantial difference has been observed for `all of the cosets' over a single one when the message state (as well as channel) dimension is $d=2$; though, this advantage demands controlling of huge number of causal orders in practice.
%We also show that the gain in classical communication rate decreases exponentially with the dimension of the message state, and increases rapidly with the increase in the number of causal orders. However, for qubit systems it saturates at $0.311$ bits per transmission, and can never reach the noiseless transmission scenario. Our analysis is mostly analytical. }

\maketitle
		
%\section{Introduction}

\section{Introduction} 
In classical information theory, it is assumed that the information carries are deterministic, and the transmission lines are used in a definite configuration in space as well as in fixed time \cite{shannon1948,cover1991}. However, physical systems obey principles of quantum theory and they offer resources which are not available 
in its classical counterpart. These unique resources can be harnessed to achieve communication protocols which are impossible in classical information theory 
\cite{Bennett_1999, PhysRevA.72.012338, PhysRevLett.94.230501}. These findings led to a complete revolution in quantum information theory \cite{nielsen2002,wilde_2017}. 
Still, quantum information theory assumes that the channels maintain a specific order in space and time. However, quantum theory allows the 
configurations where channels themselves are in superposition \cite{PhysRevLett.64.2965, PhysRevLett.91.067902}. Moreover, recently, it was realised that the superposition can exist also in the order of channels in time, in a scenario known as Indefinite Causal Order or quantum SWITCH 
\cite{Hardy_2007, Oreshkov_2012, PhysRevA.92.042124, PhysRevLett.121.090503, Oreshkov_2019}. In quantum SWITCH, the relative order of the two channels is indefinite, 
and gives rise to quantum advantages in reducing communication complexity \cite{PhysRevLett.117.100502, PhysRevLett.122.120504}, 
improving channel discrimination \cite{PhysRevA.86.040301, Frey2019} and quantum computing \cite{PhysRevA.88.022318}. 
%Moreover, quantum SWITCH has been experimentally realized \cite{procopio2015, Rubino_2017}, suggesting that the notion is not just a theoretical possibility.
Moreover, several proposals for an experimental realization of a quantum SWITCH has been actually build and tested \cite{procopio2015, Rubino_2017}, suggesting that the notion is not just a theoretical possibility.

Recently, in Ref.\cite{PhysRevLett.120.120502}, authors showed that one may achieve non-zero classical communication rates using two completely depolarising 
channels (CDCs) inserted into the quantum SWITCH, which has also been experimentally demonstrated later in \cite{goswami2018communicating}. In the same note, it is reported that using two completely entanglement 
breaking channels in SWITCH, one may achieve perfect quantum communication \cite{chiribella2018indefinite, salek2018quantum,  PhysRevLett.124.030502}. After these findings, 
several applications of quantum SWITCH have been discovered in quantum metrology, quantum thermometry and quantum information as well \cite{mukhopadhyay2019superposition, mukhopadhyay2018superposition, zhao2019quantum, 8966996, gupta2019transmitting,PhysRevA.101.012340,PhysRevA.99.062317,abbott2018communication}.

Extension of such settings beyond superposition of two channels is an immediate and interesting generalization to make to see whether it provides %more 
bigger communication advantage. 
However, such generalization comes with a serious concern whether it is not out of the experimental scope. 
In Ref.\cite{PhysRevA.101.012346}, authors showed that there is almost twofold increase in communication rate %gain 
if causal superposition of $3$ channels is used instead of $2$ channel causal superposition. On the other hand, 
the number of relevant configurations jumps from $2$ to $3!=6$. This makes experimental implementation very cumbersome, 
but nevertheless, possible. Furthermore, their numerical results suggests that usage of $3$ channels in three cyclic causal orders gives %almost 
similar gain as all $3!$ causal orders. 
This bolsters the idea that considering $N$ causal orders for $N$ channels should be efficient. An extension 
to $N$ channels with $N!$ causal orders has been proffered in Ref.\cite{e21101012}, however, they used numerical approach to find the communication rates which might %cause 
suffer from numerical errors. An analytical approach is in demand to delve deeper into these matters and to answer the following open questions: a) Can $N$ channels in quantum SWITCH 
allow perfect transmission of classical information?, and b) Can we achieve substantial gain in classical communication rates with optimal number of causal orders in a quantum SWITCH? We answer these questions in detail in this paper.

Cyclic permutations of $N$ elements form a subgroup of all $N$ element permutations. 
In this paper, we find that all cosets of permutation group factorized with respect to cyclic permutations behave equivalently, when they determine used casual order superposition of CDCs in a quantum SWITCH. Therefore, we consider single coset or multiple coset causal superposition. We refer to causal orders of channels from a single coset as cyclic causal orders, and we term superposition of causal orders of channels from more than one coset as non-cyclic. 
We analytically find classical communication capacity for $N$ CDCs inserted into a quantum SWITCH, which superposes $M\in [2,N]$ cyclic causal orders. Similarly, we derive some results for $M \in [2,N!]$ non-cyclic causal orders, when $N=3$ and $N=4$.  
We find that for the cyclic case classical communication rate depends on $M\leq N$, the number of superposed cyclic permutations, but does not depend on $N$, the number of CDCs in the quantum SWITCH. If we keep on increasing $M$ (and necessarily increasing also $N$), we observed that the communication rate increases rapidly with the increase in the number of causal orders, but it never reaches noiseless transmission. For example, it saturates at $0.311$bits for qubit systems. 
For non-cyclic case, the increase of the classical communication rate is not directly linked to $M$. 
On the other hand, we uncover that the classical communication rate decreases almost exponentially with the dimension of the message state, which seems a bit counter-intuitive at first sight. 

The rest of the paper is organized as follows. In the next section, we will briefly introduce the  
quantum Switch formalism. In Sec.\ref{secti-1} we present  
results for $N$ completely depolarizing channels in a quantum Switch, while superposing only cyclic causal orders. Sec.\ref{sectio-3} contains the detailed analysis of $N$ CDCs with arbitrary non-cyclic causal orders in a quantum SWITCH with special emphasis on $N=3$ and $4$. Finally, we conclude in Sec.\ref{sectio-2}.

\section{Quantum SWITCH and quantum channels in superposition of different causal orders}
Quantum communication devices can be modelled as quantum channels, i.e., a completely positive and trace preserving linear maps, $\Lambda: {\mathcal L}({\mathcal H})\to {\mathcal L}({\mathcal H})$. Any such map admits Kraus decomposition, i.e., $\Lambda(\rho)=\sum_i K_i\rho K_i^\dagger$, where $\{K_i\}$ is a set of Kraus operators with $\sum_iK_i^\dagger K_i={\mathbb{I}}$, and $\rho\in {\mathcal L}({\mathcal H})$.

In this work, we are considering a scenario where $N$-channels are put into a coherent superposition of their differently ordered concatenations. Originally, quantum SWITCH was used to construct a superposition of $N=2$ causal orders \cite{PhysRevA.88.022318}. In this case, quantum SWITCH is a higher order map which takes two channels as input, and outputs the 
superposition of their orders based on the state of the control qubit.  Mathematically, quantum SWITCH transforms two input channels $\Lambda_1$ and $\Lambda_2$, with Kraus decomposition $\{K_i^{(1)}\}$ and $\{K_i^{(2)}\}$ respectively, into the overall channel 
\begin{align*}
 %S(\Lambda_1,\Lambda_2)(\star)=\sum_{ij}W_{ij}(\star)W_{ij}^{\dagger},
 S(\Lambda_1,\Lambda_2)(\cdot)=\sum_{ij}W_{ij}(\cdot)W_{ij}^{\dagger},
\end{align*}
whose Kraus operators $W_{ij}$ are defined as 
\begin{align*}
W_{ij}=\ketbra{0}{0}\otimes K_i^{(1)} K_j^{(2)}+\ketbra{1}{1}\otimes K_j^{(2)} K_i^{(1)}.
\end{align*}
Note that though $W_{ij}$ depends on the specific Kraus decomposition of channels $\Lambda_1$ and $\Lambda_2$, the effective quantum channel $S(\Lambda_1,\Lambda_2)$ depends only on the input channels, allowing SWITCH to be a valid %quantum map 
higher order map 
\cite{PhysRevA.88.022318}. 
We can extend the SWITCH formalism for more than two inputs, i.e., for $N>2$ \cite{e21101012,COLNAGHI20122940}. In this case, the extended SWITCH is a higher order map which takes $N$ channels as input, and outputs the superposition of orders based on the state of the control system
that must have sufficiently high dimensionality. 
Then for $N$-channels, $\{\Lambda_p\}$, with Kraus representations $\{K_j^{(p)}\}$, the extended SWITCH will output an effective channel of the form
\begin{align}
 S(\Lambda_1,\Lambda_2,..,\Lambda_N)(\cdot)=\sum_{ij\cdots \eta} W_{ij\cdots \eta}(\cdot) W_{ij\cdots \eta}^{\dagger}\hspace{0.2cm},
\end{align}
whose Kraus operators $W_{ij\cdots \eta}$ are defined as
\begin{equation}
\begin{split}
W_{ij\cdots \eta} = \sum_{\ell=0}^{M-1}  \vert \ell \rangle\langle \ell \vert \otimes \mathcal{P}_{\ell} (K_i^{(1)}, K_j^{(2)},\cdots, K_{\eta}^{(N)}), 
\end{split}
\label{n-mKraus}
\end{equation}
where $M\in [2,N!]$ and $\mathcal{P}_{\ell} \in {\bf S}_N$ represents concatenation of $N$ operators reordered according to the permutation $j$, e. g., $\mathcal{P}_{0} (K_i^{(1)}, K_j^{(2)},\cdots, K_{\eta}^{(N)})=K_i^{(1)} K_j^{(2)}\cdots K_{\eta}^{(N)}$. For brevity, we will drop the upper index `$(p)$' in the rest of the paper.%in the following texts.
 
To have a 
simple, but sufficient picture suitable for further considerations % picture, 
let us consider two unitary channels, $U_1$ and $U_2$ and the control qubit in the state $\ket{\psi}_c=\frac{1}{\sqrt{2}}(\ket{0}+\ket{1})$.  For the pure state $\ket{\Psi}$ of the message system (we often call it a target state as well), the output of the SWITCH will be a pure state
$S(U_1,U_2)(\ket{\psi}_c \bra{\psi}\otimes \ket{\Psi} \bra{\Psi})=\ket{\xi} \bra{\xi}$, 
where 
$\ket{\xi}=\frac{1}{\sqrt{2}}(\ket{0}\otimes U_1U_2\ket{\Psi}+\ket{1}\otimes U_2U_1\ket{\Psi})$. 
To see the interference phenomenon, one needs to measure the control qubit in the 
Fourier basis, i.e., $\{\ket{\pm}=\frac{1}{\sqrt{2}}(\ket{0}\pm\ket{1})\}$, then the 
resulting target state will take the form $\ket{\Psi_{\pm}^f}=\frac{1}{\sqrt{2}}( U_1U_2\pm U_2U_1)\ket{\Psi}$.
\begin{figure}[htb]
\centering
\includegraphics[scale=0.13]{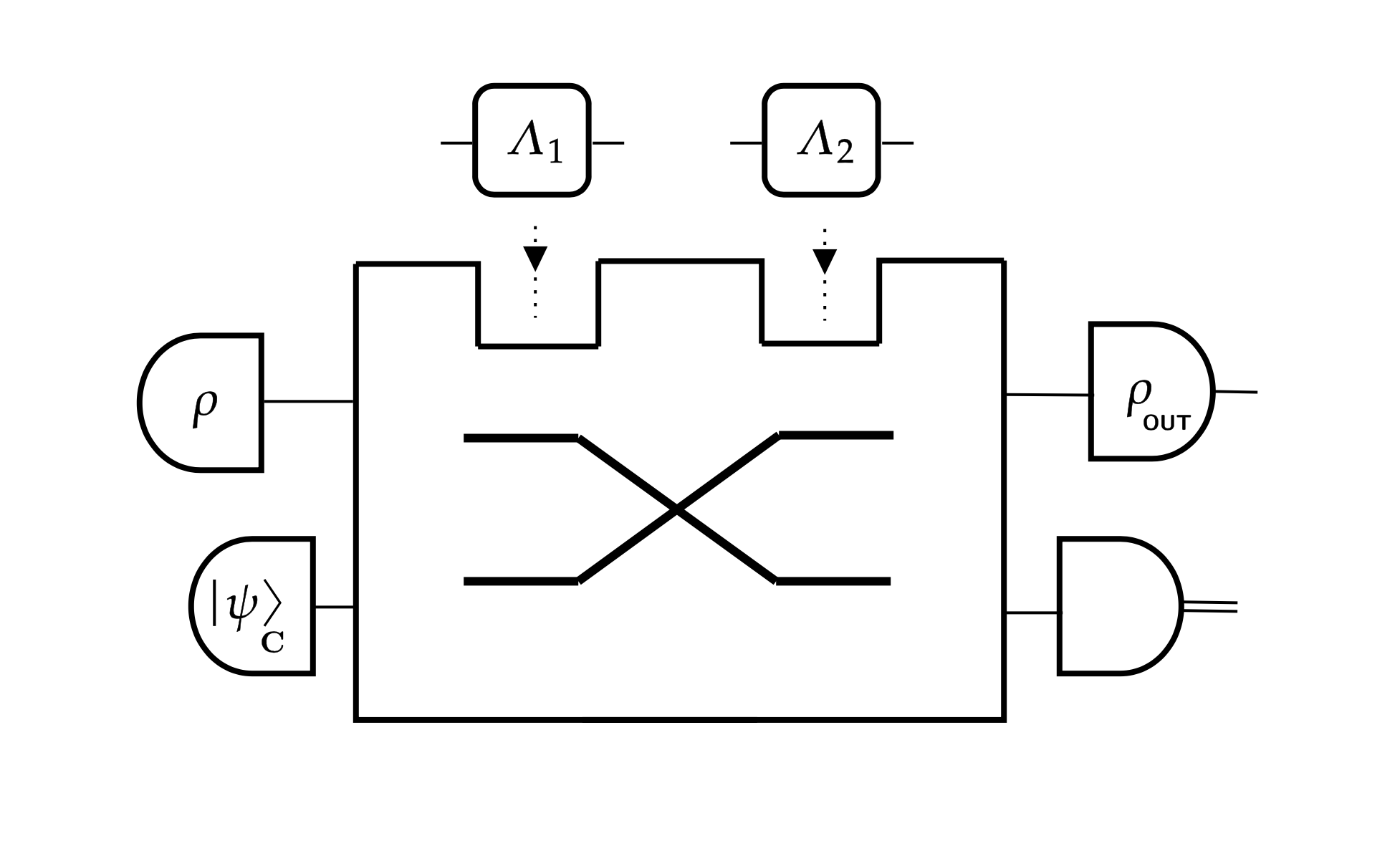}
\caption{ \textit{Illustration of a Quantum SWITCH}.--  Based on the state of the control qubit, $\ket{\psi}_c$, SWITCH takes two channels $\Lambda_1$ and $\Lambda_2$ as input, and outputs: a) either %of the definite orders 
$\Lambda_1\circ \Lambda_2$ or $\Lambda_2\circ \Lambda_1$ if control qubit is in $\ket{0}$ or $\ket{1}$, respectively, b) a superposition of causal orders if $\ket{\psi}_c=\frac{1}{\sqrt{2}}(\ket{0}+\ket{1})$. Here the message quantum state is $\rho$ and $\rho_{out}$ is the final output after post-selecting the control qubit.}
\label{fig:d}
\end{figure}

Extending this idea to $N$ channels is possible \cite{e21101012}, and the number of possible causal orders increases to $M\in[2,N!]$. 
%For the sake of simplicity, one can consider
For the brevity of explanation, let's consider 
 $N$ unitary channels $\{U_j\}$ and the control state $\frac{1}{\sqrt{M}}\sum_{j=0}^{M-1}\ket{j}$, then the final target state after the measurement of the control system will be 
$\ket{\Psi_k^f}=\frac{1}{\sqrt{M}}(\sum_{j=0}^{M-1} \langle e_k\ket{j}\mathcal{P}_j(U_1,U_2,...,U_{N})\ket{\Psi}$, 
where $\{\ket{e_k};\ket{e_k}=\frac{1}{\sqrt{M}}\sum_{j=0}^{M-1}e^{2\mi\pi jk}\ket{j}\}$ is the Fourier basis of $\{\ket{j}\}$.

%%%%%%%%%%%%%%%%%%%%%%%%%%%%%%%%%%%%%%%%%%%%%%%%%%%%%%%%%%%%%%%%%%%%%%%%%%%%%%%%%%%%%%%%%
\section{Using cyclic orders of $N$ completely depolarizing channels in quantum SWITCH}
\label{secti-1}
%%%%%%%%%%%%%%%%%%%%%%%%%%%%%%%%%%%%%%%%%%%%%%%%%%%%%%%%%%%%%%%%%%%%%%%%%%%%%%%%%%%%%%%%%%%

A completely depolarizing channel in $d$-dimensions can be described by 
%\begin{align}
%    \label{eq:defCDC}
%    \Lambda(\rho)=\frac{1}{d^2}\sum_{i=1}^{d^2}U_i\rho U_i^{\dagger}=\frac{1}{d}{\rm Tr}[\rho]\mathbb{I}_d,
%\end{align}
\begin{align}
    \label{eq:defCDC}
    \Lambda(X)=\frac{1}{d^2}\sum_{i=1}^{d^2}U_i X U_i^{\dagger}=\frac{1}{d}{\rm Tr}[X]\mathbb{I}_d,
\end{align}
where $\{U_i;\:\:i=1,2,...,d^2\}$ are $d\times d$ unitary operators satisfying ${\rm Tr}[U_i^\dagger U_j]=d\delta_{ij}$, $\mathbb{I}_d$ is identity operator of order $d$, %and $\rho$ is an arbitrary $d$-dimensional density matrix. 
and $X$ is an arbitrary linear operator on $d$-dimensional Hilbert space. 
Direct transmission of information through single or several concatenated %completely depolarizing channels 
CDCs 
necessarily results in zero classical communication rate. In contrast, it was shown that given two identical 
CDCs labeled as $\Lambda_1$ and $\Lambda_2$ and a control qubit state, $\ket{\psi}_c=\frac{1}{\sqrt{2}}(\ket{0}+\ket{1})$, 
%where $\ket{\psi}=\sum_{i=0}^{1} c_i\ket{i}$ with $c_i\in \mathcal{R}^+$, 
there is a possibility of non-zero classical communication using quantum SWITCH \cite{PhysRevLett.120.120502}.

Here, we will generalize the scheme represented in \cite{PhysRevLett.120.120502} to $N$ CDCs $\{\Lambda_i\}$. We will be considering first only %$N$ 
$M\in[2,N]$ possible cyclic orders. Accordingly, the state of control qubit is $\ket{\tilde{\psi}}_c=\frac{1}{\sqrt{M}}\sum_{j=0}^{M-1}\ket{j}$. %, where $M\in[2,N]$. 
Therefore, the Kraus operators of a channel resulting from $N$ CDCs in extended quantum SWITCH %higher order map 
 can be written as (see Eq.(\ref{n-mKraus}))
\begin{equation}
\begin{split}
K_{ij..\eta} =\frac{1}{d^{N} } \sum_{\ell=0}^{M-1}  \vert \ell \rangle\langle \ell \vert \otimes \mathcal{P}_{\ell}^{(c)} (U_i, U_j,.., U_{\eta}), \vspace{0.3cm} ,
\end{split}
\end{equation}
where $\mathcal{P}_{\ell}^{(c)} (U_i, U_j,.., U_{\eta})$ defines the cyclic permutations of unitaries. For example, for  $N=3$ the cyclic permutations are $\mathcal{P}_{0}^{(c)} (U_i, U_j,U_k)=U_i U_j U_k$, $\mathcal{P}_{1}^{(c)} (U_i, U_j,U_k)=U_j U_k U_i$ and $\mathcal{P}_{2}^{(c)} (U_i, U_j,U_k)=U_k U_i U_j$. 
If the sender prepared the target system in the state $\rho$, then the receiver will receive the output from the Quantum SWITCH as
\begin{align}
\label{eq:rhomgen}
 &\rho_{M}  := S(\Lambda_1,\Lambda_2,..,\Lambda_N)(\rho_c\otimes\rho)   \nonumber\\
 & = \frac{1}{M d^{2N}}\sum_{i,j,\ldots,\eta} \sum_{l=0}^{M-1} \sum_{l'=0}^{M-1} \ket{l}\bra{l'}\otimes \nonumber\\
 & \quad \otimes
 {P}_{l}^{(c)} (U_i, U_j,.., U_\eta)\; \rho\: \left({P}_{l'}^{(c)} (U_i, U_j,.., U_\eta)\right)^{\dagger}
\end{align}
where $\rho_c=\ketbra{\tilde{\psi}}{\tilde{\psi}}_c$. 
For cyclic permutations of orders we will see below that only two types of contributions are present in the final output state: i) diagonals ($l=l'$) which are proportional to $\mathbb{I}$ and ii) the off-diagonals ($l\neq l'$) which are proportional to $\rho$. %It is easy to show that
All the diagonal terms are equivalent to the following prototype form
\begin{align}
 &\frac{1}{d^{2N}}\sum_{ij\cdots \eta=1}^{d^2} \overbrace{U_{\eta}\cdots U_{j}}^{N-1}  U_{i}\left( \rho \right) U_{i}^{\dagger} \overbrace{U_j^{\dagger}\cdots U_{\eta}^{\dagger}}^{N-1} \nonumber\\
 &=\frac{1}{d^{2N-2}}Tr(\rho)\sum_{j\cdots \eta=1}^{d^2} \overbrace{U_{\eta}\cdots U_{j}}^{N-1} \frac{\mathbb{I}}{d} \overbrace{U_j^{\dagger}\cdots U_{\eta}^{\dagger}}^{N-1} \nonumber\\
 &=\frac{1}{d^{2N-2}} d^{2(N-1)}\frac{\mathbb{I}}{d}=\frac{\mathbb{I}}{d},
\end{align}
and all off-diagonal terms are equivalent to the term below
\begin{align}
\label{eq:offdiagterms1}
 &\frac{1}{d^{2N}}\sum_{ij\cdots \eta=1}^{d^2}  \overbrace{U_{\eta}\cdots U_{\mu}}^k U_{\ell}\cdots U_{j} U_{i}  \left( \rho \overbrace{U_{\mu}^{\dagger}\cdots U_{\eta}^{\dagger}}^k \right) U_{i}^{\dagger} U_{j}^{\dagger}\cdots U_{\ell}^{\dagger} \nonumber\\
 & = \frac{1}{d^{2N-2}}\sum_{j\cdots \eta=1}^{d^2} Tr\left( \rho \overbrace{U_{\mu}^{\dagger}\cdots U_{\eta}^{\dagger}}^k \right)  \overbrace{U_{\eta}\cdots U_{\mu}}^k U_{\ell}\cdots U_{j} \frac{\mathbb{I}}{d} U_{j}^{\dagger}\cdots U_{\ell}^{\dagger} \nonumber\\
 & = \frac{d^{2(N-k-1)}}{d^{2N-2}}\sum_{\mu\cdots \eta=1}^{d^2} Tr\left( \left(\rho \overbrace{U_{\mu}^{\dagger}\cdots }^{k-1}\right) \frac{1}{\sqrt{d}} U_{\eta}^{\dagger}\right)  \frac{1}{\sqrt{d}}U_{\eta}\left( \overbrace{\cdots U_{\mu}}^{k-1}\right)   \nonumber\\
  & =\frac{1}{d^{2k}} d^{2(k-1)} \rho=\frac{\rho}{d^2},
\end{align}
where $k\in \{1,\ldots,N-1\}$, we used $\sum_{i=1}^{d^2} \frac{1}{d} Tr(X U_i^\dagger) U_i= X$ and Eq. (\ref{eq:defCDC}) multiple times. Note that the index $k$ represents all possible cyclic $k$-shifts.  %  permutations. 
Therefore, for $N$ CDCs in SWITCH  the final output state for $M\in [2,N]$ causal orders %with quantum SWITCH 
is 
\begin{equation}
\rho_M= \frac{1}{M}\left(\mathbb{I}\otimes \frac{\mathbb{I}}{d}+\sum_{i\neq j}\vert i \rangle\langle j \vert \otimes \frac{\rho}{d^2}\right).
\label{n-nmat}
\end{equation}
We note that the output density matrix does not depend on $N$ and only its off-diagonal entries depend on $\rho$. %(cf. Appendix.\ref{block-ndnd}).
%Mario's suggestion
Further, if the control system is measured in the Fourier basis and the outcome is known, the target system will regain dependence on $\rho$ and one can extract information about it \cite{PhysRevLett.120.120502}. Therefore, there is a possibility of nonzero classical communication according to HSW theorem \cite{651037,PhysRevA.56.131}. 
%Now, one can observe that the information is encoded into the correlations, which will be lost if we trace out the control system or measure it in the computational basis.  
In what follows, we will quantitatively investigate the above scheme.
%Clearly, the output state depends on the input state. Therefore, there is a possibility of nonzero classical communication according to HSW theorem \cite{651037,PhysRevA.56.131}. Now, one can observe 
%that the information is encoded into the correlations, which will be lost if we trace out the control system or measure it in the computational basis.  However, if the control system is measured in the Fourier basis, the target system will regain dependence on $\rho$ and one can extract information about it \cite{PhysRevLett.120.120502}.
%Now we will quantitatively investigate our scheme. 

\subsection{ Method} 
%Mario's suggestions
Classical capacity of quantum communication channel $\Lambda$ is characterized by the Holevo quantity, which is defined as 
%Classical communication rate of a channel is directly proportional to the Holevo quantity. The Holevo quantity of a quantum channel, 
%$\Lambda$ is defined as $\chi(\Lambda)=\max_{\{p_i\rho_i\}} I(A:B)_{\varrho}$, where $\varrho=\sum_ip_i\ketbra{i}{i}\otimes \Lambda(\rho_i)$ with input target state as $\rho=\sum_ip_i\rho_i$. Therefore, in terms of von Neumann entropy, the mutual information, $I(A:B)_{\varrho}=H(\Lambda(\rho))-\sum_ip_iH(\Lambda(\rho_i))$, which yields
\begin{align}
    \chi(\Lambda)=\max_{\{p_i\rho_i\}}\left[H(\Lambda(\rho))-\sum_ip_iH(\Lambda(\rho_i))\right]. %,
\end{align}
%where $\Lambda$ is the quantum channel. 
In the Ref.\cite{PhysRevLett.120.120502}, authors found how to evaluate Holevo quantity for two quantum channels when the information is send through a pair of quantum channels processed by a quantum SWITCH.
%In the Ref.\cite{PhysRevLett.120.120502}, authors found a practical way how to evaluate Holevo quantity for two quantum channels in a quantum SWITCH. 
Since, this method works with the channel induced by the SWITCH on the control plus target system after the insertion of the depolarizing channels, it automatically work also for $N$-channels in the generalized SWITCH.  

Therefore, the Holevo quantity of $N$ CDCs in the SWITCH is given by (we refer readers to the supplementary material of Ref.\cite{PhysRevLett.120.120502})
\begin{align}
    \chi^{(M)}=\log d+H\left(\tilde{\rho_c}(M)\right)-H_{\min}(\rho_M), \vspace{0.3cm} \forall M\in[2,N];
    \label{mainchi}
\end{align}
where $\tilde{\rho_c}(M)$ is reduced state of control qubit after evolution (see Eq. (\ref{n-nmat})), i.e., $\tilde{\rho_c}(M)=\frac{1}{M}(\sum_i\ketbra{i}{i}+\frac{1}{d^2}\sum_{i\neq j}\ketbra{i}{j})$ and $H_{\min}$ is min-entropy, i.e., $H_{\min}(\rho_M)=\min_{\rho} H(\rho_M)$ with $H(.)$ being the von Neumann entropy. However, the main difficulty will be to calculate the eigenvalues of the $Md\times Md$ matrix, $\rho_M$ to evaluate $H_{\min}(\rho_M)$. Fortunately, we are able to use the method given in Ref. \cite{10.2307/3620776,powell2011calculating} to find its eigenvalues. The determinant of the matrix, $\rho_M$ 
%with $c_i=\frac{1}{\sqrt{N}}$ 
is given by 
\begin{align}
 {\rm Det}(\rho_M)   
=& {\rm Det}\Big(\frac{1}{M}\Big[\frac{\mathbb{I}}{d} -\frac{\rho}{d^2}\Big]\Big)^{\times (M-1)} \nonumber\\ &\times {\rm Det}\Big(\frac{1}{M}\Big[\frac{\mathbb{I}}{d} + (M-1)\frac{\rho}{d^2}\Big]\Big),
\end{align} 
where ${\rm Det}(.)^{\times (M-1)} $ denotes that there are $M-1$ products of same determinant (Full details of the calculations are given in Appendix \ref{block-ndnd}). This beautiful simplified form of determinant of actual $Md\times Md$ matrix tells us that finding the eigenvalues of actual matrix has reduced to finding the eigenvalues of these small matrices, i.e., $\frac{1}{M}\Big[\frac{\mathbb{I}}{d} -\frac{\rho}{d^2}\Big]$ with degeneracy $M-1$ and $\frac{1}{M}\Big[\frac{\mathbb{I}}{d} + (M-1)\frac{\rho}{d^2}\Big]$ with degeneracy one. As $[\mathbb{I},\rho]=0$, the eigenvalues of the matrix $\rho_M$ will be the union of the eigenvalues of these two smaller matrices with their appropriate degeneracy. Let $\{\lambda^+_i\}_{i=1}^d$ and $\{\lambda^-_i\}_{i=1}^d$ be the eigenvalues of $\frac{1}{M}\Big[\frac{\mathbb{I}}{d} + (M-1)\frac{\rho}{d^2}\Big]$ and $\frac{1}{M}\Big[\frac{\mathbb{I}}{d} -\frac{\rho}{d^2}\Big]$ respectively, then 
\begin{align*}
    \lambda_i^+=\frac{1}{Md}+\frac{M-1}{Md^2}\lambda_i^\rho,\:\:\mbox{and}\:\:
    \lambda_i^-=\frac{1}{Md}-\frac{1}{Md^2}\lambda_i^\rho,
\end{align*}
where $\{\lambda_i^\rho\}_{i=1}^d$ are eigenvalues of $\rho$. As $H_{\min}(\rho_M)=\min_{\rho} H(\rho_M)$, certainly the minima will be ascertained if $\lambda_i^{\rho}=1$ and $\lambda_j^{\rho}=0$ with $i\neq j$. Therefore, with the constrain that $\sum_i\lambda_i^{\rho}=1$, we can find that
\begin{align}
    H_{\min}(\rho_M) = -\Big\{\frac{d+(M-1)}{Md^2}\log\frac{d+(M-1)}{Md^2}\:\:\:\:\:\:\:\:\:\nonumber\\+\frac{(M-1)(d-1)}{Md^2}\log\frac{(d-1)}{Md^2} 
+\frac{d-1}{d}\log\frac{1}{Md}\Big\}.
\label{minEntro}
\end{align}
Now, the remaining task is to find the expression for $H(\tilde{\rho_c}(M))$, which is given by
\begin{align}
    H(\tilde{\rho_c}(M))=-\Big(\frac{M-1+d^2}{Md^2}\log \frac{M-1+d}{Md^2}\nonumber\\+(M-1)\frac{d^2-1}{Md^2}\log \frac{d^2-1}{Md^2}\Big).
    \label{redEntro}
\end{align}
With these expressions, we can evaluate the classical communication rate, $\chi^{(M)}$ for $N$ CDS with SWITCH from Eq.(\ref{mainchi}) for cyclic causal orders $M\in [2,N]$. Notice that for $M=2$, it reduces to the result for $N=2$ scenario as discussed in Ref.\cite{PhysRevLett.120.120502}. This observation tells us that the gain in classical communication depends only on the number of superposed causal orders, $M$.

\vspace{0.2cm}
\subsection{ Results} 
\begin{figure}[htb]
	\centering
	\includegraphics[scale=0.65]{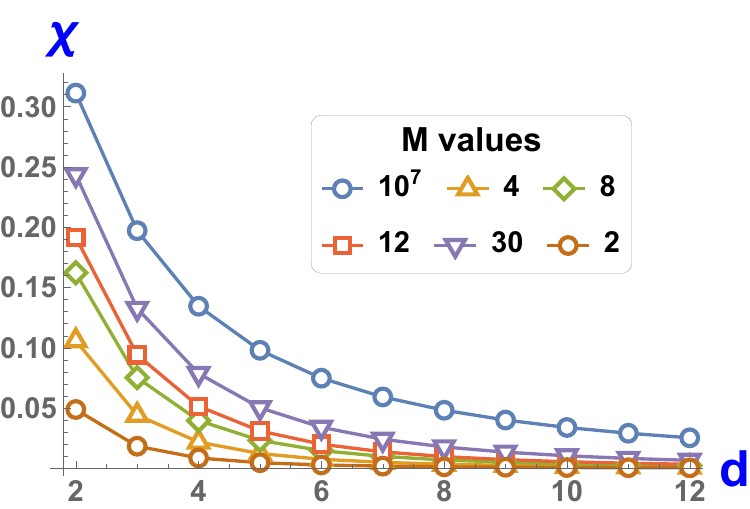}
	\caption{(Color online) Plot illustrates that the Holevo quantity for different number of causal orders $M$ is almost-exponentially decreasing with the dimension $d$ of the target state $\rho$.}
	\label{fig:a}
\end{figure}

\begin{figure}[htb]
	\centering
	\includegraphics[scale=0.56]{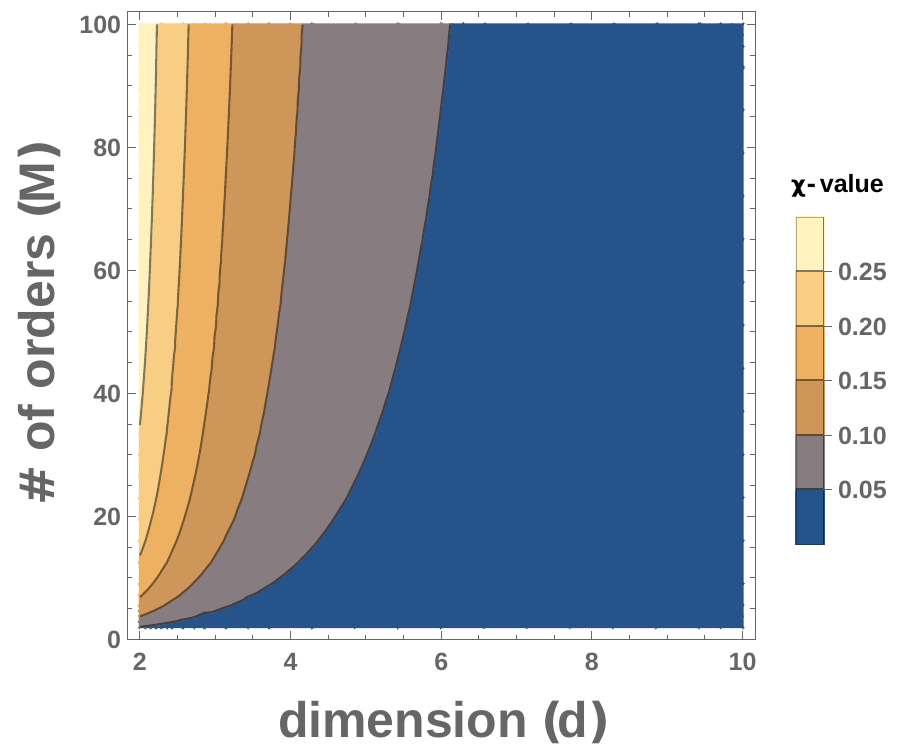}
	\caption{(Color online) Contour plot depicts the Holevo quantity for different number of causal orders $M$ and the dimension $d$ of the target state $\rho$. It shows that %the communication advantages is shifting towards
	higher classical communication rates are achieved for lower $d$ and higher $M$.}
	\label{fig:b}
\end{figure}

\begin{figure}[htb]
	\centering
	\includegraphics[scale=0.45]{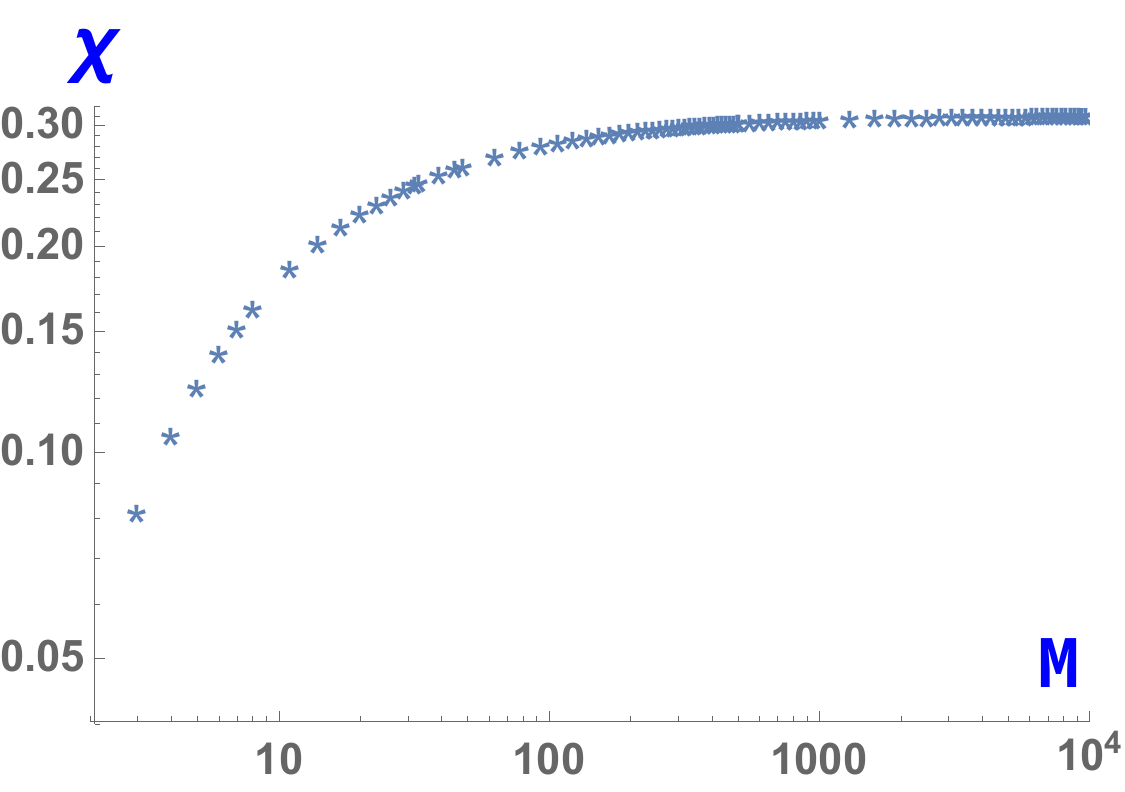}
	\caption{(Color online) Log plot depicts that the Holevo quantity is increasing for different number of causal orders $M$ for $d=2$. It shows that the communication rate is converging to the value $0.311$ bits per single system transmission.}
	\label{fig:c}
\end{figure}

%To have a better understanding, we plot the Holevo quantity $\chi^{(M)}$ with respect to dimension ($d$) of the input state $\rho$ as well as with the number of causal orders. 
In order to illustrate the behaviour of the Holevo quantity $\chi^{(M)}$ with respect to dimension, $d$, of the input state $\rho$ and the number of causal orders we prepared a plot in Fig. \ref{fig:a}. We 
find that the classical communication capacity increases %the noise reduction increases 
as we increase $M$, however, it decreases almost-exponentially as $d$ increases. %(see Fig.\ref{fig:a}). 
Fig.\ref{fig:b} shows the communication rates for different choice of ($M,d$). It is clear from the contour plot that the higher values of communication rates are achieved with smaller $d$ values as well as higher $M$ values. This means that using quantum SWITCH with $M$ causal orders in $d=2$  will offer maximum classical communication rate. However, we find that the communication rate %advantage 
saturates with the increase of $M$, %the number of causal orders, 
hinting that it is not possible to reach perfect communication %rate 
in the asymptotic limit, i.e., $\chi^{(\infty)}\neq 1$ (see Fig.\ref{fig:c}). To prove this claim, we write down the expression for Holevo quantity for $d=2$, i.e., 
\begin{align*}
    \chi^{(M)}|_{d=2}=1+\frac{1}{4}\left[\log_2\frac{4}{27}+(1+\frac{1}{M})\log_2(1+\frac{1}{M})\right.\\ \left.+\frac{1}{M}\log_2 \frac{27}{M^2}-(1+\frac{3}{M})\log_2(1+\frac{3}{M})\right].
\end{align*}
Therefore, at $M \to \infty$, the Holevo quantity, $\chi^{(\infty)}|_{d=2}\cong 1+\frac{1}{4}\log_2\frac{4}{27}\approx 0.311$bits/transmission.


%\section{Generalization to $M$ $\in [2,N!]$ causal orders}
\section{Generalization to various combinations of cyclic and non-cyclic causal orders}\label{sectio-3}
%\section{ $N$ completely depolarizing channels with SWITCH in non-cyclic orders}
%As $N$ elements can have $N!$ possible permutations, there exists $N!$ causal orders of $N$ channels. 
There are $N!$ possible permutations of $N$ elements, thus there exists $N!$ causal orders of $N$ channels. 
However, above, we have considered only $M\leq N$ cyclic causal orders of channels, 
%because -- (i) the increment of communication advantage is substantial to notice, and (ii) it becomes really cumbersome to implement superposition of $N!$ causal orders in an experiment.
because the increase of communication rate is substantial to notice and it becomes problematic to implement superposition of $N!$ causal orders in an experiment.

In this section, we will discuss whether the extension from $N$ cyclic causal orders to all $N!$ orders for $N$ CDCs is 
provides significant improvement %worth the gain 
in the communication rates. In a recent work Ref.\cite{e21101012}, authors showed that communication rates do not increase %linearly
evenly with the increase of the number of causal orders. %It is because mixing of cyclic and non-cyclic orders hinders the benefit (see Appendix.\ref{did_not}). 
We think that this behavior can be atributed to inevitable mixing of cyclic and non-cyclic causal orders, which hinders the potential benefit due to appearance of input state independent terms (see Appendix \ref{did_not}).
This is why we considered the cyclic orders separately. 
 
By considering any order of action of $N$ channels as the zeroth element and then performing cyclic permutations of this element, we can form a coset of cyclic causal orders (coset of permutation group with respect to the subgroup of cyclic permutations of $N$ elements) which contains $N$ elements. In this way, we can identify $\frac{N!}{N}=(N-1)!$ such cosets in the set of all $N!$ permutations. 
We refer to these individual cosets as to cyclic causal orders, and each of them yields the same classical communication rate, which we already evaluated in the previous section.

However, if we consider superposition of orders of elements from different cosets the situation becomes a bit demanding. 
%For the cyclic case, the off-diagonal contributions are equal to $\frac{\rho}{d^2}$ (see Eq. (\ref{eq:offdiagterms1})), however, this is no longer the case if we consider all possible permutations. 
Off-diagonal contributions to $\rho_M$ that map between control system states belonging to a single coset act on the message system as  $\frac{\rho}{d^2}$ (see Eq. (\ref{eq:offdiagterms1})), however, this is no longer the case if the control system states are not from the same coset. 
%we consider any pair of possible permutations, i.e. pair control system states

Let us note that as the number of channels increases the number of types of off-diagonals contributions increases and will include %at least 
the terms presented in %according to 
the following Table \ref{table:A1}. 
For derivation of the terms appearing in the table we refer the reader to Appendix \ref{did_not}. 
In accordance with Table \ref{table:A1} %shows that 
for $N\geq 3$ we found that lot of cross-coset off-diagonal terms are proportional to $\mathbb{I}$ %as well as
and the terms which are proportional to $\rho$ have a decreasing weight factor as $N$ increases. 
To illustrate the impact of the above findings we present a case study for $N=3$ and $N=4$.
%\begin{widetext}
\begin{table}[!ht]
\large
\centering
\begin{tabular}{|c|c|}\hline
	  & Terms in off-diagonal block  \\ \hline
	$N=2$ & $ \frac{\rho}{d^2}$ \\ \hline
	$N=3$ &  $\frac{\rho}{d^2}$, $\frac{\mathbb{I}}{d^3}$\\ \hline
	$N=4$ & $\frac{\rho}{d^2}$, $\frac{\mathbb{I}}{d^3}$, $\frac{\rho}{d^4}$ \\ \hline
	$\vdots$ & $\vdots$ \\ \hline
	$N=2k$ & $\frac{\rho}{d^2}$, $\frac{\mathbb{I}}{d^3}$, $\frac{\rho}{d^4}$, $\cdots$, $\frac{\mathbb{I}}{d^{N-1}}$, 
	$\frac{\rho}{d^N}$ \\ \hline
	$N=2k+1$ & $\frac{\rho}{d^2}$, $\frac{\mathbb{I}}{d^3}$, $\frac{\rho}{d^4}$, $\cdots$, $\frac{\rho}{d^{N-1}}$, 
	$\frac{\mathbb{I}}{d^N}$ \\ \hline
\end{tabular}
\caption{Entries for off-diagonal blocks in the output density matrix for $N$ channels in the SWITCH when we are considering all possible causal orders, i.e., $M\in [2,N!]$. 
%(for proof, we refer readers to Appendix.\ref{did_not})
Note that off-diagonal terms within a single coset are always $\frac{\rho}{d^2}$ (see, Appendix \ref{did_not})}.
\label{table:A1}
\end{table}
%\end{widetext}

%%%%%%%%%%%%%%%%%%%%%%%%%%%%%%%%%

\subsection{Case study for $N=3$}
%\subsection{N=3 case}
%\subsection{Three channels in a SWITCH}
For three CDCs, there will be two cosets of cyclic causal orders. 
The Kraus operator for three CDCs with SWITCH can be expressed as
% \begin{equation}
% \begin{split}
$d^3 K_{ijk} = \sum_{\ell=0}^{M-1}  \vert \ell \rangle\langle \ell \vert \otimes \mathcal{P}_{\ell}(U_i, U_j,U_k)$, 
% \end{split}
% \end{equation}
where $2\leq M \leq 6$. For the message qubit prepared in $\rho$, the evolved state at the output of quantum SWITCH is $\rho_{M_1,M_2}=S(\Lambda_1, \Lambda_2, \Lambda_3)(\rho_c\otimes\rho)$, where $M_1,M_2$ denotes the number of causal orders from two cosets respectively, and $M_1+M_2=M$. 
%We find that the expression for $\rho_{M_1,M_2}$ is given by
In appendix \ref{3channAll} we find that 
%the expression for $\rho_{M_1,M_2}$ is given by
\begin{align}
%\begin{split}
\rho_{M_1,M_2}=\frac{1}{M}\left\{\mathbb{I}_c\otimes \frac{\mathbb{I}}{d}+L_{M_1,M_2}\otimes \frac{\rho}{d^2}+B_{M_1,M_2}\otimes \frac{\mathbb{I}}{d^3}\right\}, 
%\end{split}    
\end{align}
where $2\leq M \leq 6$. The matrix $L$ and $B$ are $M\times M$ matrices with the following form,
\begin{align*}
{\bf L}=\left(
  \begin{array}{cc}
    S_{M_1\times M_1} & {\bf 0}_{M_1\times M_2}\\
    {\bf 0}_{M_2\times M_1} & S_{M_2\times M_2}\\
  \end{array}
\right),\:\:
{\bf B}=\left(
  \begin{array}{cc}
    {\bf 0}_{M_1\times M_1} & {\bf 1}_{M_1\times M_2}\\
    {\bf 1}_{M_2\times M_1} & {\bf 0}_{M_2\times M_2}\\
  \end{array}
\right),
\end{align*}
where the matrix $S=\sum_{i\neq j=0}^{m-1}\ketbra{i}{j}$, ${\bf 0}$ is a null matrix and ${\bf 1}$ is a matrix with all entries equal to one. 
%every element one.
For such scenario, the general expression for the classical communication for three CDCs with $M\in [2, 6]$ causal orders 
can efficiently be written as before
\begin{align}
    \chi^{(M_1,M_2)}=\log d+H(\tilde{\rho}_c(M))-H_{\min}(\rho_{M_1,M_2}),
    \label{mainchi:Ne3}
\end{align}
where $\tilde{\rho}_c(M)$ is reduced state of the control %qubit 
system after evolution, i.e., $\tilde{\rho}_c(M)=\frac{1}{M}(\mathbb{I}+\frac{1}{d^2}\sum_{i\neq j}\ketbra{i}{j})$.
%and $H_{\min}$ is min-entropy.

To evaluate $H(\tilde{\rho}_c(M))$ and $H_{\min}(\rho_{M_1,M_2})$, we need to diagonalize the matrices, $\tilde{\rho}_c(M)$ and $\rho_{M_1,M_2}$ respectively. We know that the expression for the $H(\tilde{\rho}_c(M))$ is given in Eq.(\ref{redEntro}). However, diagonalizing $\rho_{M_1,M_2}$ is %another story
much more complicated. 
%It is very had (not impossible!) to diagonalize the matrix using the method given in Appendix.\ref{diag-bloch-pres}. We find that for $M_1=M_2$, $[L,M]=0$ and therefore, the matrices $L$ and $B$ are simultaneously diagonalizable (for detailed calculations, we refer to Appendix.\ref{3channAll}).
Analytical diagonalization is done in Appendix \ref{3channAll} using the method in Appendix \ref{diag-bloch-pres}. 
%We performed these tedious calculations and placed them and the related details in the Appendix \ref{3channAll}.
%Analytical diagonalization is possible using the method given in Appendix \ref{diag-bloch-pres}. We performed these tedious calculations and placed them and the related details in the Appendix \ref{3channAll}. 
%However, we will keep the focus of the main text just on the obtained results. In particular, we find 
It follows that for $M_1=M_2$, $[L,B]=0$ and therefore, the matrices $L$ and $B$ are simultaneously diagonalizable. 
%Then, for $M_1=M_2$, the expression for $H_{\min}(\rho_{M_1,M_1})$ is given by
Consequently, $H_{\min}(\rho_{M_1,M_1})$ is given by
\begin{align*}
 -H_{\min}(\rho_{M_1,M_1})=\lambda_+\log\lambda_++\lambda_-\log\lambda_-~~~~~~~~~~~~~~~~ \\+(d-1)\left[\lambda_+^0\log\lambda_+^0 +\lambda_-^0\log\lambda_-^0\right.\nonumber\\ \left.+(M-2)\left\{\frac{1}{Md^2}\log\frac{(d-1)}{Md^2} +\frac{1}{Md}\log\frac{1}{Md}\right\}\right],
\end{align*}
where $\lambda_{\pm}=\frac{d^2+(M_1-1)d\pm M_1}{Md^3}$ and $\lambda_{\pm}^0=\frac{d^2\pm M_1}{Md^3}$ with $M=2M_1$. We show in Appendix \ref{3channAll} that the output states for $M_1\neq M_2$ can also be diagonalized using the method presented in Appendix \ref{diag-bloch-pres}.

To further elucidate our findings here, we compute the Holevo quantity for all $(M_1,M_2)$ values in the Table \ref{tabole1} for $d=2$. We find that the Holevo quantity is higher for the case when either $M_1=0$ or $M_2=0$ compared to $M_1\neq M_2$. However, we find that the Holevo quantity reaches its maximum for $M_1=M_2=3$.
% 
% Furthermore, we find that our Table.\ref{tabole1} shows more classical communication 
% rates than that of Table.I of Ref.\cite{PhysRevA.101.012346}. This observation reflects that there might be a problem in numerical method employed in Ref.\cite{PhysRevA.101.012346}.
\begin{table}[ht]
\centering
\begin{tabular}{|l|c|c|c|c|}
\hline
\diagbox[dir=NW]{\rule{0mm}{0mm}$M_1$}{$M_2$} & 0 & 1 & 2 & 3 \\
\hline
\:\:\: 0                     & - & 0 & 0.0488 & 0.0817 \\
\hline
\:\:\: 1                        & 0 & 0 & 0.0334 & 0.0640\\
\hline
\:\:\: 2                     & 0.0488 & 0.0334 & 0.0524 & 0.0767 \\
\hline
\:\:\: 3                      & 0.0817 & 0.0640 & 0.0767 & 0.0981 \\
\hline
\end{tabular}
\label{tab:Pugh_matrix2}
\caption{Table shows the communication rates $\chi^{(M_1,M_2)}$ for three CDCs %with 
in superposition of $M=M_1+M_2$ causal orders from two cosets.
% with possible ($M_1, M_2$) values such that $M_1+M_2=M$. 
Here, we consider message state to be a qubit ($d=2$). Table shows that the Holevo quantity is %always more 
higher for the cases when either $M_1$ or $M_2$ is zero.}
\label{tabole1}
\end{table}

\begin{figure}[htb]
	\centering
	\includegraphics[scale=0.65]{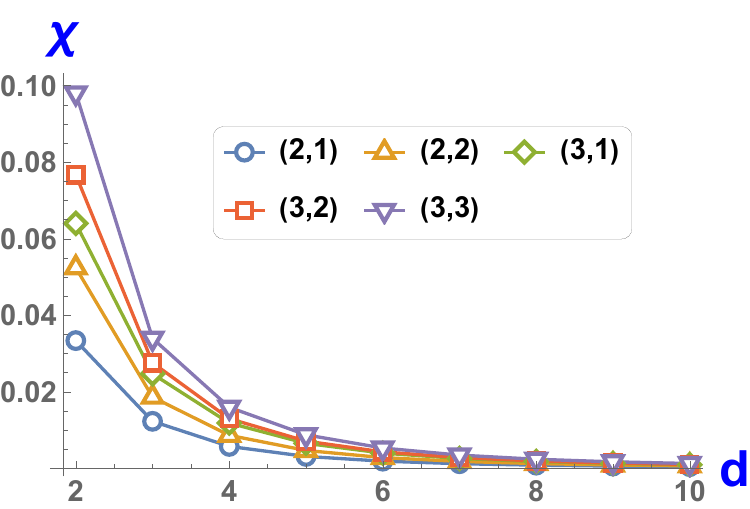}
	%\caption{(Color online) Plot illustrates that the Holevo quantity of different $(M_1,M_2)$ values for $N=3$ is almost-exponentially %decreasing with the dimension $d$ of the target state, $\rho$.}
	\caption{(Color online) Holevo quantity for $N=3$ and different $(M_1,M_2)$ values depending on the dimension $d$ of the target state.}
	
	\label{fig:d}
\end{figure}

\begin{figure}[htb]
	\centering
	\includegraphics[scale=0.65]{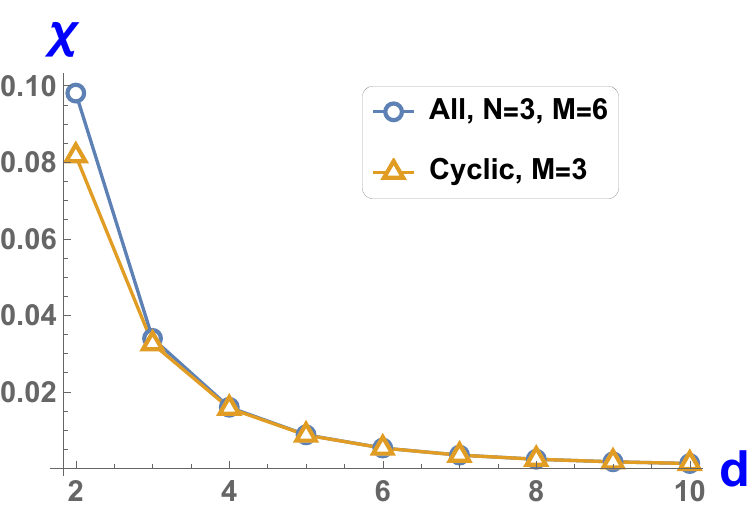}
	%\caption{(Color online) The plot of the Holevo quantity for %the case of 
	%\tb{$(N=3, M=6)$ (all cosets) as well as for $(N=3, M=3)$ (cyclic) against} the dimension $d$ of the target state, $\rho$. It shows %that both options provide the same classical communication rates except for $d=2$ where $(N=3, M=6)$ dominates.}
	\caption{(Color online) Holevo quantity as a function of target state dimension $d$ for $(N=3, M=6)$ , $(N=3, M=3)$.}
	\label{fig:e}
\end{figure}

We also plot the Holevo quantity for different $(M_1, M_2)$ values with respect to the dimension of the target state $d$ in Fig.\ref{fig:d}, and find that the Holevo quantity decreases with $d$ in an exponential-like fashion. In Fig.\ref{fig:e}, we plot the Holevo quantity of $(N=3,M=6)$ (all cosets) as well as $(N=3,M=3)$ (cyclic orders) against the dimension $d$. The figure shows that both scenarios yield the same communication rates except for $d=2$, where the former dominates. This %unique 
plot indicates that it might be %sufficient 
efficient to consider only cyclic permutations (one coset).

%%%%%%%%%%%%%%%%%%%%%%%%%%%%%%%%%%%%%%%%

\subsection{Case study for $N=4$}
For four channels there are six ($(4-1)!=6$) cosets of cyclic causal orders. Let us denote $M_{\eta}$ as the number of causal orders in a coset where $M_{\eta}\in[0,3]$ for each $\eta\in[1,6]$. Let us consider that the target state is $\rho$. Now, if we consider $M = (\sum_{\eta}M_{\eta})$ causal orders in a quantum SWITCH, the output state will have -- a) within each cyclic coset, the diagonal terms are proportional to $\frac{\mathbb{I}}{d}$ and off-diagonal terms are proportional to $\frac{\rho}{d^2}$; and  b) cross-coset off-diagonal terms are proportional to $\frac{\mathbb{I}}{d^3}$ as well as $\frac{\rho}{d^4}$ (for detailed calculation see Appendix \ref{did_not}).
We will use a particular way of listing causal orders for each coset. Namely, the first entry of each coset is linked with %the zeroth
the order $\Lambda_1\Lambda_2\Lambda_3\Lambda_4$ by some %non-cyclic 
permutation of the last three labels. In this way, the first element in each coset is starting with $\Lambda_1$. And rest is %will be
constructed using cyclic permutations from first element of corresponding coset. 
For $\eta=1,2,...,6$, the zeroth elements are respectively $\Lambda_1\Lambda_2\Lambda_3\Lambda_4$,  $\Lambda_1\Lambda_2\Lambda_4\Lambda_3$, $\Lambda_1\Lambda_3\Lambda_2\Lambda_4$, $\Lambda_1\Lambda_3\Lambda_4\Lambda_2$, $\Lambda_1\Lambda_4\Lambda_2\Lambda_3$ and $\Lambda_1\Lambda_4\Lambda_3\Lambda_2$.
Using this setup, we find that the output state after evolution driven by the SWITCH is given by
\begin{align}
 %\rho_{M_{\eta}}=\frac{1}{M}\left\{\mathbb{I}_c\otimes\frac{\mathbb{I}}{d}+ L_{M_{\eta}}\otimes \frac{\rho}{d^2}+ B_{M_{\eta}}\otimes \frac{\mathbb{I}}{d^3} +Q_{M_{\eta}}\otimes \frac{\rho}{d^4}   \right\},
\rho_{M_{\eta}}=\frac{1}{M}\left\{\mathbb{I}_c\otimes\frac{\mathbb{I}}{d}+  \left(\frac{L_{M_{\eta}}}{d^2}+\frac{Q_{M_{\eta}}}{d^4}\right)\otimes \rho+ B_{M_{\eta}}\otimes \frac{\mathbb{I}}{d^3}    \right\}, 
 \label{4chanellsA}
\end{align}
where the matrices, $L$, $B$ and $Q$ are specified in Appendix \ref{did_not}. Also, in this case, the general expression for the classical communication for four CDCs with $2\leq M\leq 24$ causal orders 
can efficiently be written as 
\begin{align}
    \chi^{(M_{\eta})}=\log d+H(\tilde{\rho}_c(M_{\eta}))-H_{\min}(\rho_{M_{\eta}}),
    \label{mainchi:Ne4}
\end{align}
where $\tilde{\rho}_c(M_{\eta})$ is reduced state of control qubit after evolution, i.e., $\tilde{\rho}_c(M_{\eta})=\frac{1}{M}(\mathbb{I}+\frac{1}{d^2}\{L_{M_{\eta}}+B_{M_{\eta}} \}+\frac{1}{d^4}Q_{M_{\eta}})$.

It is very hard to evaluate Holevo quantity for arbitrary $\{M_{\eta}\}$, %values, 
as diagonalizing $\rho_{M_{\eta}}$ is usually hard %(not impossible!) 
analytically. Therefore, we mostly resort to numerical %routes 
approach. However, there are specific cases where it is possible to analytically diagonalize $\rho_{M_{\eta}}$, e.g., the scenario with $(M_3=M_5=4;M=8)$ (We refer readers to Appendix \ref{4chanState} for complete analysis). 

\begin{figure}[htb]
	\centering
	\includegraphics[scale=0.65]{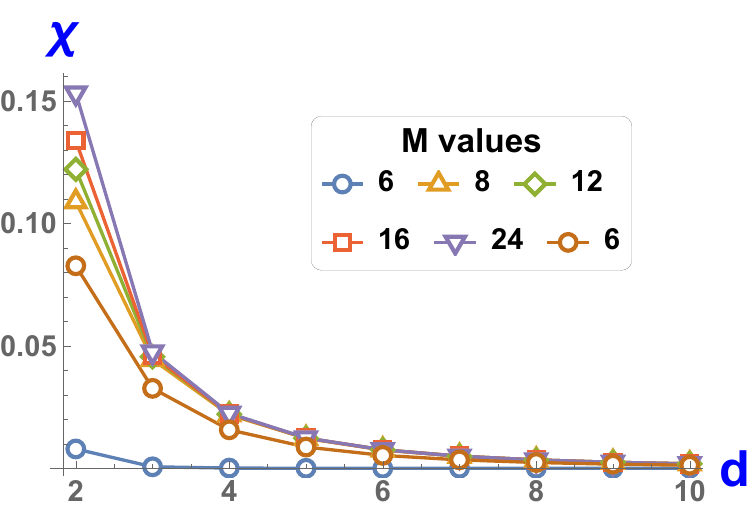}
	%\caption{(Color online) Plot illustrates that the Holevo quantity for different values of $M$ for $N=4$ is almost-exponentially decreasing with the dimension $d$ of the target state, $\rho$.}
	\caption{(Color online) Holevo quantity for $N=4$ and different $M_\eta$ values depending on the dimension $d$ of the target state. Please see the main text for detailed description of the curves. }
	\label{fig:f}
\end{figure}

\begin{figure}[htb]
	\centering
	\includegraphics[scale=0.65]{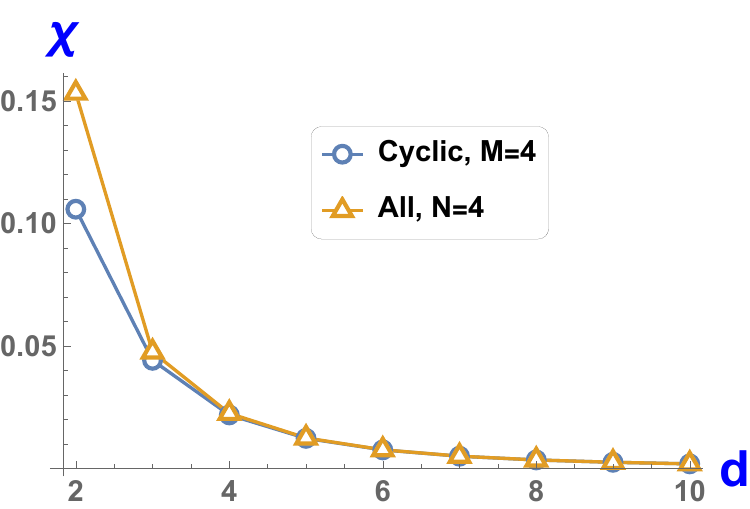}
	%\caption{(Color online) The plot of the Holevo quantity for the case of $(N=4, M=24)$ (all cosets) as well as cyclic, $M=4$ against the dimension $d$ of the target state, $\rho$. It shows that both yield same amount of classical communication rates except for $d=2$ where $(N=4, M=24)$ dominates.}
	\caption{(Color online) Holevo quantity as a function of target state dimension $d$ for $(N=4, M=24)$ , $(N=4, M=4)$.}
	\label{fig:g}
\end{figure}

%To illucidate our findings here, 
To graphically illustrate our findings for $N=4$ we present two plots.
in Fig.\ref{fig:f} %-- a) 
we plot %maximally achievable 
dependence of Holevo quantity on $d$ for different $M$ values, i.e., %6^{\times 2}
$M=\{6, 8, 12, 16, 24\}$. For $M=6$, we consider two scenarios -- a) $(M_1=4, M_2=2)$ and b) $M_{\eta}=1\forall \eta$ (see red and blue lines (with circular points) respectively in Fig.\ref{fig:f}). In former case, we consider a situation such that only two cross-coset off-diagonal terms are dependent on $\rho$. However, in later case, we consider maximum number of $\rho$ dependent cross-coset off-diagonal terms (see Appendix \ref{4chanState}). Fig.\ref{fig:f} shows 
that the Holevo quantity is decreasing in exponential-like fashion with $d$. Unlike Fig.\ref{fig:d}, here the Holevo quantity for %$M=8(M_3=4,M_5=4)$, $M=12(M_1=4,M_3=4,M_5=4)$, $M=16(M_1=4,M_3=4,M_5=4,M_6=4)$, and $M=24$ 
all the plotted $M_\eta$ 
are very close to each other except for $d=2$. 
%the case for $d=2$. 
This is due to the presence of a big number of cross-coset off-diagonal terms $\frac{\rho}{d^4}$ in these scenarios. 
In Fig.\ref{fig:g} we plot the Holevo quantity for $(N=4, M=24)$ and its cyclic counter part $(N=4, M=4)$ with respect to message system dimension $d$. 
%In Fig.\ref{fig:g}, we plot the Holevo quantity of $(N=4,M=24)$ (all cosets) as well as cyclic orders, $M=4$ against the dimension $d$.
Analogously to Fig. \ref{fig:e}, also here the plot shows that both scenarios provide the same communication rates except for $d=2$. 
%This unique plot also indicates that it might be sufficient to consider cyclic permutations (one coset) only. 
%Also, we find an interesting finding from two Figs.\ref{fig:e} and \ref{fig:g} that the gap between Holevo quantities for all cosets vs one coset increases for $d=2$ as we increase $N$. 
%\label{sectio-2}
Comparing Figs.\ref{fig:e} and \ref{fig:g} we uncover that the gap between Holevo quantity for 
%all cosets and one coset
$(N,M=N!)$ (all cosets) and $(N,M=N)$ (one coset) increases for $d=2$ as we increase $N$ from three to four.

%%%%%%%%%%%%%%%%%%%%%%%%%%%%%%%%%%%%%%%%%%%%%%%%%%%%%
\section{Conclusions}\label{sectio-2}
%To summarize, we have studied the transmission of classical information for $N$ CDCs with arbitrary number of causal orders in a quantum SWITCH. Our approach is mostly analytical in nature. 

Completely depolarizing channel erases all information about its input state and always prepares a completely mixed state. Thus, sequential application of two such channels on the same system in any fixed order must have zero classical (or quantum) communication capacity. It was a rather surprising finding of Ebler, Salek and Chiribella \cite{PhysRevLett.120.120502} that %superposition of two causal orders created 
processing of two depolarizing channels 
by Quantum SWITCH folllowed by suitable control system measurement enables nonzero classical communication rate. 

In this paper, we studied a generalization of this scenario to $N$ completely depolarizing channels inserted into (generalized) quantum SWITCH. 
Quickly growing number of possible causal orders ($N!$) might become a roadblock for experimental realization of the SWITCH, thus one might wonder if less demanding superpositions of causal orders could provide similar advantages. As previous numerical results for $N=3$ show \cite{PhysRevA.101.012346}, already $M=N$ superposed causal orders can provide almost the fully achievable communication rate. 
We provide analytical results for transmission of classical information via superpositions of $M$ cyclically permuted Completely depolarizing channels. 
We find that the Holevo quantity is increasing with $M$ and is independent of $N$. %However, 
Surprisingly, the classical capacity decreases with the dimension $d$ of the message system. 
We found that the classical communication rate for a qubit never reaches $1$ if we are increasing $M$ (and inevitably also $N$). It saturates at around $0.311$ bits per transmission. 
%This is a non-trivial finding which suggests that quantum SWITCH with more channels may not achieve perfect classical communication capacity. 
%As we already mentioned there are $N!$ possible causal orders for $N$ channels and $N$ of these permutations for. 
Out of $N!$ possible causal orders for $N$ channels there are $N$ cyclic permutations forming a subgroup. Factoring the permutation group with respect to it we obtain $(N-1)!$ cosets, each of which is shown to be equally usable for the investigated task. However, for general $N$ we did not consider all possible causal orders together as the cross-coset off-diagonal terms are mostly independent of the message state. Instead, for $N=3,4$ we studied separately cyclic, all causal orders and various superpositions of causal orders consisting of different number of terms from different cosets. 
For the causal orders from arbitrary number of cosets, i.e., non-cyclic case, we find that with growing $N$ the cross-coset off-diagonal terms have smaller scaling factors and are either proportional to Identity or to the message state $\rho$. Therefore, we find that the Holevo quantity doesn't always increase as we increase the number of superposed causal orders. 
Our findings support the belief that considering superpositions of cyclic causal orders might yield almost optimal classical communication rates for $N$ CDCs in the SWITCH.

\vspace{1.5cm}

\textit{Acknowledgement}:-- 
SS acknowledges the financial support through the {\v S}tefan Schwarz stipend from Slovak Academy of Sciences, 
Bratislava. MS and SS acknowledge the financial support through the project OPTIQUTE (APVV-18-0518) and HOQIT (VEGA 2/0161/19). 
MS was also supported by Grant No. 61466 from the John Templeton Foundation, as part of the “The Quantum Information Structure of Spacetime (QISS)” Project (qiss.fr). The opinions expressed in this publication are those of the author(s) and do not necessarily reflect the views of the John Templeton Foundation. 
%\tb{We thank Prof. Mario Ziman for going through the manuscript and for valuable feedback.}
We thank Mario Ziman for valuable discussions. SS thanks Nidhin for a small yet crucial help.
\vspace{2cm}

{\em Note added:} After publishing first version of our work in arXiv, we noticed a similar work by Chiribella et. al. \cite{chirib_same}  appeared in arXiv which independently derives one of our results. They also claim that the classical capacity of $N$-channels with superposition of cyclic orders is exactly equal to the Holevo quantity, which will also strengthen our results in Section.\ref{secti-1}. 
\newpage

%%%%%%%%%%%%%%%%%%%%%%%
\appendix
\begin{widetext}
%%%%%%%%%%%%%%%%%%%%%%%%%%%%%%%%
\section{ Calculations for N channels with cyclic orders in a quantum SWITCH} \label{block-ndnd}
Initially, we calculated the determinant of the matrix from Eq. (\ref{n-nmat}) for small values of $M$ using the Eq. (\ref{det_formula}).  It also allowed us to anticipate its form for general $M$. However, the following lemma will be proved in a simpler way using the properties of block circulant matrix \cite{Kaveh2011}.

\begin{lemma}
For hermitian matrix $\rho$ determinant of  $Md \times Md$ matrix $\rho_M$ defined in Eq. (\ref{n-nmat}) is
 \begin{align}
 {\rm Det}(\rho_M)   
=  {\rm Det}\Big(\frac{1}{M}\Big[\frac{\mathbb{I}}{d} + (M-1)\frac{\rho}{d^2}\Big]\Big)\times  {\rm Det}\Big(\frac{1}{M}\Big[\frac{\mathbb{I}}{d} -\frac{\rho}{d^2}\Big]\Big)^{\times (M-1)},
\label{NN-deter}
\end{align} 
where ${\rm Det}(.)^{\times (M-1)} $ denotes that there are $M-1$ products of same determinant. 
%However, it is not easy to show how we reach to above simplification. The way we infer it is from the pattern that we get 
%from the cases with $N=2,3,4,....$. To convince our readers, we will give a simpler proof below using the properties of block circulant matrix \cite{Kaveh2011}.
\end{lemma}

\begin{proof}
We will begin here by mentioning some properties of block circulant matrix \cite{Kaveh2011}. A block circulant matrix ${\bf C}$ of the form 
\begin{equation*}
{\bf C}=\left(
  \begin{array}{ccccc}
    A_{0} & A_{1}  & \cdots & A_{M-2} & A_{M-1} \\
    A_{M-1}  & A_{0} &  \cdots & A_{M-3} & A_{M-2} \\
    \vdots  & \vdots & \ddots & \vdots &\vdots \\
    A_{2} & A_{3} & \cdots & A_{0}& A_{1} \\
    A_{1} & A_{2} & \cdots & A_{M-1}& A_{0} \\
  \end{array}
\right),
\end{equation*}
can be written in the following form 
\begin{align*}
{\bf C}=\sum_{i=0}^{M-1} {\bf P_i}\otimes A_i,
\end{align*}
where ${\bf P_i}$ are Permutation matrices. Now, one can Block-diagonalise ${\bf C}$ using the properties of ${\bf P_i}$ matrices as mentioned in Ref.\cite{Kaveh2011}. For the matrix $\rho_{M}$ in Eq.(\ref{n-nmat}), we can decompose it to the following form
\begin{align}
\rho_M={\bf P_0}\otimes \frac{\mathbb{I}}{Md}+\left(\sum_{i=1}^{M-1} {\bf P_i}\right)\otimes \frac{\rho}{Md^2},\nonumber\\
={\bf P_0}\otimes \frac{\mathbb{I}}{Md}+ {\bf S}\otimes \frac{\rho}{Md^2},
\label{rep-perm}
\end{align}
where ${\bf S}=\sum_{i=1}^{M-1} {\bf P_i}$ is a symmetric matrix with the entries, $ {\bf S}_{ij}=1-\delta_{ij}$, 
\begin{equation}
{\bf S}=\left(
  \begin{array}{ccccc}
    0 & 1 & 1 & \cdots & 1 \\
    1 & 0 &  1 & \cdots & 1 \\
    \vdots  & \vdots &  \ddots & \vdots & \vdots \\
    1 & 1 & \cdots & 0 & 1 \\
    1 & 1 & \cdots & 1 & 0 \\
  \end{array}
\right)
_{M\times M}.
\label{S-matrix}
\end{equation}
Therefore the block-diagonalisation of $\rho_M$ will depend on the properties of 
the matrix ${\bf S}$ alone as ${\bf P_0}=\mathbb{I}$. Then, one can define a matrix function 
$G: {\mathcal X}\to {\mathcal M}^2$ such that 
\begin{align}
G(x)=x^0 \otimes \frac{\mathbb{I}}{Md}+ x^1\otimes \frac{\rho}{Md^2},
\end{align}
where each element of an object in ${\mathcal X}$ is mapped to a $d\times d$ matrix. 
%, in general, ${\mathcal X}$ contains a set of complex numbers. Obviously, 
If $x$ is a (complex) number then the symbol `$\otimes$' will just be a product and the zeroth power of $x$ is a number one. Hence, one can show that $\rho_M=G({\bf S})$, since ${\bf S}^0=\mathbb{I}$. 
%Moreover, $[S,P_0]=0$ , we can diagonalize simultaneously. However, for shake of the proof, we know that $S^0=\mathbb{I}=P_0$ and therefore the claim.

Next, we diagonalize the matrix ${\bf S}$. Its characteristic equation %of this matrix 
is ${\rm Det}({\bf S}-\lambda \mathbb{I})=0\Rightarrow \{\lambda-(M-1)\}(\lambda+1)^{M-1}=0$, i.e, eigenvalues of ${\bf S}$ are 
$M-1$ with degeneracy one and $-1$ with degeneracy $M-1$. So, one can find an unitary, $T$ which diagonalises 
${\bf S}$, such that $T^\dagger {\bf S} T={\rm diag}(M-1, -1,-1,\cdots,-1)$. Consequently, matrix of the form 
$\tilde{T}=T\otimes \mathbb{I}$ block-diagonalises $\rho_M$ as 
\begin{align*}
\tilde{T}^\dagger \rho_M \tilde{T}={\bf P_0}\otimes \frac{\mathbb{I}}{Md}+ T^\dagger{\bf S}T\otimes \frac{\rho}{Md^2},
\end{align*}
i.e., one can write $\tilde{T}^\dagger \rho_M \tilde{T}={\rm diag}(G(M-1), G(-1),\cdots G(-1))$. Hence, one can conclude validity of Eq.(\ref{NN-deter}).
\end{proof}

Equipped with the above lemma and its proof it is an easy task to show that the characteristic equation for $\rho_M$ is of the form
\begin{align}
 {\rm Det}(\rho_M-\lambda \mathbb{I})   =0 \Rightarrow
   {\rm Det}\Big(\frac{1}{M}\Big[\frac{\mathbb{I}}{d} + (M-1)\frac{\rho}{d^2}\Big]-\lambda \mathbb{I}\Big)\times  {\rm Det}\Big(\frac{1}{M}\Big[\frac{\mathbb{I}}{d} -\frac{\rho}{d^2}\Big]-\lambda \mathbb{I}\Big)^{\times (M-1)}=0.
\end{align} 
Therefore, the eigenvalues of $\rho_M$ can be obtained as the union of eigenvalues of matrices $\frac{1}{M}\Big[\frac{\mathbb{I}}{d} + (M-1)\frac{\rho}{d^2}\Big]$ with degeneracy one and $\frac{1}{M}\Big[\frac{\mathbb{I}}{d} -\frac{\rho}{d^2}\Big]$ with degeneracy $M-1$.

\section{All possible causal orders}\label{did_not}
Here we sketch a proof for the entries in the Table \ref{table:A1}.    
Below, in Eq. \ref{3channAll}, we show that %for three 
for more than two channels, if we go beyond cyclic causal orders, the off-diagonal terms are also proportional to $\frac{\mathbb{I}}{d^3}$ along with the terms proportional to $\frac{\rho}{d^2}$. Also, in Eq. \ref{4chanState}, we show that some of the off-diagonal terms for %four channels 
more than three channels are proportional to $\frac{\rho}{d^4}$. Using these observations, we sketch how off-diagonal terms appear for arbitrary $N$. Let us begin with the following term:
% {\bf Diagonal--}
\begin{align}
  \frac{1}{d^{2N}}\sum_{ij\cdots \eta} \overbrace{U_{\eta}\cdots U_{\ell}}^{N-3}U_k U_jU_i\left(\rho U_j^\dagger U_k^\dagger \right) U_i^\dagger \overbrace{U_{\ell}^\dagger\cdots U_{\eta}^\dagger}^{N-3}
  =\frac{1}{d^{2N}}\cdot d^2\cdot d^2 \cdot d^{2(N-3)}{\rm Tr}[\rho]\frac{\mathbb{I}}{d}
   =\frac{\mathbb{I}}{d^3}.
   \label{id3}
\end{align}
Next, consider the following off-diagonal term
% {\bf Off-dagonal--}
\begin{align}
  \frac{1}{d^{2N}}\sum_{ij\cdots \eta} U_j\overbrace{U_{\eta}\cdots U_m}^{N-4} U_kU_{\ell} U_i\left(\rho U_j^\dagger U_k^\dagger U_{\ell}^\dagger \right) U_i^\dagger \overbrace{U_m^\dagger\cdots U_{\eta}^\dagger}^{N-4}
   =\frac{1}{d^{2N}}\cdot d^2\cdot d^2 \cdot d^{2(N-4)} \rho =\frac{\mathbb{\rho}}{d^4}.
   \label{id4}
\end{align}
It is evident from above calculation that the term $\frac{\mathbb{\rho}}{d^4}$ will not occur if $N< 4$. Now, notice the similarity between two Eqs.(\ref{id3}, \ref{id4}). They are quite similar except the positioning of $U_j$. Using this connection, we consider the following term 
\begin{align*}
  \frac{1}{d^{2N}}\sum_{ij\cdots \eta} \overbrace{U_{\eta}\cdots U_n}^{N-5}U_j U_m U_k U_{\ell}U_i\left(\rho U_j^\dagger  U_m^\dagger U_k^\dagger U_{\ell}^\dagger \right) U_i^\dagger \overbrace{U_n^\dagger\cdots U_{\eta}^\dagger}^{N-5}
  =\frac{1}{d^{2N}}\cdot d^2\cdot d^2\cdot d^2 \cdot d^{2(N-5)}{\rm Tr}[\rho]\frac{\mathbb{I}}{d}
   =\frac{\mathbb{I}}{d^5}.
\end{align*}
The above term will not occur when $N<5$. Therefore, we can infer the terms proportional to $\frac{\mathbb{I}}{d^{2k}}$ and $\frac{\mathbb{\rho}}{d^{2k}}$, where $k\in \mathbb{Z}^+$;
\begin{align*}
  \frac{1}{d^{2N}}\sum_{ij\cdots \eta} \overbrace{U_{\eta}\cdots U_{\mu}}^{N-(2k+1)}U_j \overbrace{U_{p}\cdots U_k}^{2(k-1)} U_{\ell}U_i\left(\rho U_j^\dagger  \overbrace{U_{p}^\dagger\cdots U_k^\dagger}^{2(k-1)} U_{\ell}^\dagger \right) U_i^\dagger \overbrace{U_{\mu}^\dagger\cdots U_{\eta}^\dagger}^{N-(2k+1)}
 =\frac{1}{d^{2N}}\cdot d^4\cdot d^{2(k-1)}d^{2[N-(2k+1)]}{\rm Tr}[\rho]\frac{\mathbb{I}}{d}
   =\frac{\mathbb{I}}{d^{2k+1}}.
\end{align*}
Again, we can comment that the above term will not occur if $N <(2k+1)$. 
%And the final term 
Analogously, we have the term 
\begin{align*}
  \frac{1}{d^{2N}}\sum_{ij\cdots \eta} U_j\overbrace{U_{\eta}\cdots U_{\mu}}^{N-2k} \overbrace{U_{p}\cdots U_k}^{2(k-1)}U_{\ell} U_i\left(\rho U_j^\dagger \overbrace{U_{p}^\dagger\cdots U_k^\dagger}^{2(k-1)} U_{\ell}^\dagger \right) U_i^\dagger \overbrace{U_{\mu}^\dagger\cdots U_{\eta}^\dagger}^{N-2k}
   =\frac{1}{d^{2N}}\cdot d^2\cdot d^{2(k-1)} \cdot d^{2(N-2k)} \rho =\frac{\mathbb{\rho}}{d^{2k}}.
\end{align*}
%Therefore, we show that the entries in Table.\ref{table:A1} are consistent.
Thus, we showed existence of the terms presented in Table \ref{table:A1}.

%\subsection{Case study for $N=3$}
\subsection{Three channels in a SWITCH}
%\subsection{N=3 case}
\label{3channAll}
For three CDCs, there are two cosets ($6/3=2$) of cyclic orders. Let us consider the following permutation of channels 
%as the order zeroth order, say,  
$\Lambda_2\Lambda_1\Lambda_3$. Now applying cyclic permutations to it, %operators, 
we get a coset $\{\Lambda_2\Lambda_1\Lambda_3,  \Lambda_1\Lambda_3\Lambda_2, \Lambda_3\Lambda_2\Lambda_1\}$. 
One can also see that rest of the channel orders forms the other coset. The remaining task is to see how a message state behaves when it is sent through the superposition of %all six 
$M_1+M_2=M$ causal orders. It is easy to see that off diagonal elements (in the sense of Eq.(\ref{eq:rhomgen})) within a single    %each cyclic 
coset will contribute %similarly 
in the same way as in Eq.(\ref{n-nmat}).
Next, we need to investigate the off-diagonal terms between two cosets. By analogous calculation as in Eq.(\ref{eq:offdiagterms1}), which we summarize below, we found that these are proportional to identity, implying that they will not contribute to the classical communication. 
In particular, there are two types of terms that can be evaluated as follows:
%We found two prototype combinations of such off-diagonal terms and show how they simplifies:  
\begin{align*}
 \frac{1}{d^6}\sum_{ijk} U_iU_jU_k\rho U_j^\dagger U_k^\dagger U_i^\dagger
 =\frac{1}{d^4}\sum_{ik}U_i\left(\frac{1}{d^2}\sum_j U_j(U_k\rho) U_j^\dagger\right) U_k^\dagger U_i^\dagger
   =\frac{1}{d^4}\sum_{ik}U_i\left({\rm Tr}[\rho U_k] \frac{\mathbb{I}}{d}\right) U_k^\dagger U_i^\dagger
   =\frac{1}{d^4}\sum_{i} U_i\rho U_i^\dagger =\frac{\mathbb{I}}{d^3},
\end{align*}
%where we have used the fact that $\frac{1}{d}\sum_{k}\left({\rm Tr}[\rho U_k] \right) U_k^\dagger=\rho$. And other off-diagonal term,
\begin{align*}
 \frac{1}{d^6}\sum_{ijk} U_iU_jU_k\rho U_i^\dagger U_j^\dagger U_k^\dagger
 =\frac{1}{d^4}\sum_{ik}U_i\left(\frac{1}{d^2}\sum_j U_j(U_k\rho U_i^\dagger) U_j^\dagger\right) U_k^\dagger 
   =\frac{1}{d^4}\sum_{ik}U_i\left({\rm Tr}\Big[U_i^\dagger(U_k\rho )\Big] \frac{\mathbb{I}}{d}\right) U_k^\dagger 
   =\frac{1}{d^4}\sum_{k} U_k\rho U_k^\dagger =\frac{\mathbb{I}}{d^3},
\end{align*}
%where we have used the fact that $\frac{1}{d}\sum_{i}\left({\rm Tr}\Big[U_i^\dagger( U_k\rho)\Big] \right) U_i=U_k\rho$. 
where we used $\frac{1}{d}\sum_{k}\left({\rm Tr}[\rho U_k] \right) U_k^\dagger=\rho$ and $\frac{1}{d}\sum_{i}\left({\rm Tr}\Big[U_i^\dagger( U_k\rho)\Big] \right) U_i=U_k\rho$.  
%Therefore, these terms does not contribute to the gain in classical communication rates. This is the why we consider one of cosets as cyclic orders above. 

Hence, the final output state is 
\begin{align}
% %\begin{split}
\rho_{M_1,M_2}= \frac{1}{M}\left\{\mathbb{I}_c\otimes \frac{\mathbb{I}}{d} + \left(\sum_{\ell\neq \ell'=0}^{M_1-1} \vert \ell \rangle\langle \ell' \vert_c+\sum_{\mu\neq \mu'=M_1}^{M-1} \vert \mu \rangle\langle \mu' \vert_c\right) \otimes \frac{\rho}{d^2}+\sum_{\ell,\mu}\left(\vert \ell \rangle\langle \mu \vert_c +\vert \mu \rangle\langle \ell \vert_c\right)\otimes\frac{\mathbb{I}}{d^3}\right\}.
% \nonumber\\
% =&\frac{1}{M}\left(
%   \begin{array}{cc}
%     S_{M'\times M'} & X_{M'\times (M-M')} \\
%     X^T & S_{(M-M')\times M-M')}\\
%   \end{array}
% \right)
% _{M\times M},
\label{all-orders}
% %\end{split}    
\end{align}
To find the eigenvalues of the matrix in Eq.(\ref{all-orders}) for general $M_1$ and $M_2$, %values, 
we will re-write it 
%the matrix 
in the Block form,
i.e.,
\begin{align}
 \rho_{M_1,M_2}=\frac{1}{M}\left\{\mathbb{I}_c\otimes \frac{\mathbb{I}}{d}+L_{M_1,M_2}\otimes \frac{\rho}{d^2}+B_{M_1,M_2}\otimes \frac{\mathbb{I}}{d^3}\right\},
 \label{N3M6}
\end{align}
%where the matrixes $L$ and $B$ are $M\times M$ matrix with the following form,
where $M\times M$ matrices $L$ and $B$ have the following form

\begin{align*}
{\bf L}=\left(
  \begin{array}{cc}
    S_{M_1\times M_1} & {\bf 0}_{M_1\times M_2}\\
    {\bf 0}_{M_2\times M_1} & S_{M_2\times M_2}\\
  \end{array}
\right)
_{M\times M},\hspace{1cm}\mbox{and}\hspace{1cm}
{\bf B}=\left(
  \begin{array}{cc}
    {\bf 0}_{M_1\times M_1} & {\bf 1}_{M_1\times M_2}\\
    {\bf 1}_{M_2\times M_1} & {\bf 0}_{M_2\times M_2}\\
  \end{array}
\right)
_{M\times M},
\end{align*}
where the matrix $S$ is defined in Eq.(\ref{S-matrix}), ${\bf 0}$ is a null matrix and ${\bf 1}$ is a matrix with all entries are one. For  $M_1=M_2$ $[L,B]=0$, %one can show that 
so these two matrices are simultaneously diagonalizable using a $M\otimes M$ unitary matrix, $U$, i.e., $U^\dagger L U={\rm Diag}(M_1-1, M_1-1, -1,\cdots, -1)$ %and $U^\dagger B U={\rm Diag}(\pm M_1, 0,\cdots, 0)$. 
and $U^\dagger B U={\rm Diag}(-M_1, M_1, 0,\cdots, 0)$.
Consequently, the unitary matrix $\tilde{U}=U\otimes \mathbb{I}$ will Block-diagonalize the matrix $\rho^{M_1,M_1}$, i.e., 
\begin{align}
 \tilde{U}^\dagger \rho_{M_1,M_1}\tilde{U} =\frac{1}{M}\left\{\mathbb{I}_c\otimes \frac{\mathbb{I}}{d}+U^\dagger L U\otimes \frac{\rho}{d^2}+U^\dagger B U\otimes \frac{\mathbb{I}}{d^3}\right\}.
\end{align}
This enables us to find the characteristic equations for above Block diagonal matrix as, 
%\begin{align}
 %{\rm Det}(\rho_{M_1,M_1}-\lambda \mathbb{I})   =0 \Rightarrow
   %{\rm Det}\Big(\frac{1}{M}\Big[\frac{\mathbb{I}}{d} + (M_1-1)\frac{\rho}{d^2}\pm M_1 \frac{\mathbb{I}}{d^3}\Big]-\lambda %\mathbb{I}\Big)\times  {\rm Det}\Big(\frac{1}{M}\Big[\frac{\mathbb{I}}{d} -\frac{\rho}{d^2}\Big]-\lambda \mathbb{I}\Big)^{\times %(M-2)}=0.
%\end{align}
\begin{align}
 {\rm Det}(\rho_{M_1,M_1}-\lambda \mathbb{I})   =0 \Rightarrow
   &{\rm Det}\Big(\frac{1}{M}\Big[\frac{\mathbb{I}}{d} + (M_1-1)\frac{\rho}{d^2}+ M_1 \frac{\mathbb{I}}{d^3}\Big]-\lambda \mathbb{I}\Big) \nonumber \\ %\times   \\
    &\times {\rm Det}\Big(\frac{1}{M}\Big[\frac{\mathbb{I}}{d} + (M_1-1)\frac{\rho}{d^2}- M_1 \frac{\mathbb{I}}{d^3}\Big]-\lambda \mathbb{I}\Big)\times
   {\rm Det}\Big(\frac{1}{M}\Big[\frac{\mathbb{I}}{d} -\frac{\rho}{d^2}\Big]-\lambda \mathbb{I}\Big)^{\times (M-2)}=0.
\end{align}
Therefore, the eigenvalues of $\rho_{M_1,M_1}$ is the union of eigenvalues of $\frac{1}{M}\Big[\frac{\mathbb{I}}{d} + (M_1-1)\frac{\rho}{d^2}\pm M_1 \frac{\mathbb{I}}{d^3}\Big]$ with degeneracy one and $\frac{1}{M}\Big[\frac{\mathbb{I}}{d} -\frac{\rho}{d^2}\Big]$ with degeneracy $M-2$. 

However, for $M_1\neq M_2$, we will resort to the %usual 
technique %prescribed 
described in Appendix \ref{diag-bloch-pres}. In this scenario, we have three distinct cases, i.e., $(M_1=2,M_2=1)$, $(M_1=3,M_2=1)$ and $(M_1=3,M_2=2)$. To diagonalize the output matrix $\rho_{M_1,M_2}$ for these cases, we consider the characteristic equation, ${\rm Det}(\rho_{M_1,M_2}-\lambda \mathbb{I})   =0$ and solve it to find eigenvalues. 

%%%%%%%%%%%%%%%%%%%%%%%%%%%%%%%%%%%%%%%%%%
\textbullet{\bf A1. $(2,1)$}: The characteristic equation for this case can be written using Appendix (\ref{diag-bloch-pres}), 
\begin{align}
{\rm Det}\left(\frac{1}{3}\left[\frac{\mathbb{I}}{d}-\frac{\rho}{d^2}\right]-\lambda \mathbb{I}\right)\times {\rm Det}\left(\frac{1}{9}\left[\frac{a^2\mathbb{I}}{d^2}+\frac{a\rho}{d^3}-\frac{2\mathbb{I}}{d^6}\right]\right)=0,
\end{align}
where $a=1-3d\lambda$. 
\begin{proof}
 Using Eq.(\ref{det_formula}), the characteristic determinant reduces to 
 ${\rm Det}(\rho_{2,1}-\lambda \mathbb{I})={\rm Det}\left[{\bf X}_{11}^{(2)}\right]{\rm Det}\left[{\bf X}_{22}^{(1)}\right]{\rm Det}\left[{\bf X}_{33}^{(0)}\right]$. 
 Again using Eq.(\ref{another-det}), we find ${\bf X}_{11}^{(2)}={\bf X}_{11}^{(1)}-{\bf X}_{12}^{(1)}\left({\bf X}_{22}^{(1)}\right)^{-1}{\bf X}_{21}^{(1)}$, 
 then ${\rm Det}\left[{\bf X}_{11}^{(2)}\right]{\rm Det}\left[{\bf X}_{22}^{(1)}\right]
 ={\rm Det}\left[{\bf X}_{11}^{(1)}{\bf X}_{22}^{(1)}-{\bf X}_{12}^{(1)}{\bf X}_{21}^{(1)}\right]$. 
 Now, we notice that ${\bf X}_{11}^{(1)}={\bf X}_{22}^{(1)}$ and ${\bf X}_{12}^{(1)}={\bf X}_{21}^{(1)}$ which means 
 \begin{align*}
  {\rm Det}\left[{\bf X}_{11}^{(2)}\right]{\rm Det}\left[{\bf X}_{22}^{(1)}\right]={\rm Det}\left[{\bf X}_{11}^{(1)}-{\bf X}_{12}^{(1)}\right]{\rm Det}\left[{\bf X}_{11}^{(1)}+{\bf X}_{12}^{(1)}\right].
 \end{align*}
 At this stage we will consider the product of determinants, ${\rm Det}\left[{\bf X}_{11}^{(1)}+{\bf X}_{12}^{(1)}\right]{\rm Det}\left[{\bf X}_{33}^{(0)}\right]$. Applying Eq.(\ref{det_formula}), we can simplify it as following 
 \begin{align*}
  {\rm Det}\left[{\bf X}_{11}^{(1)}+{\bf X}_{12}^{(1)}\right]{\rm Det}\left[{\bf X}_{33}^{(0)}\right]={\rm Det}\left[\left\{{\bf X}_{11}^{(0)}\right\}^2+{\bf X}_{12}^{(0)}{\bf X}_{11}^{(0)}-2\left\{{\bf X}_{13}^{(0)}\right\}^2\right],
 \end{align*}
 where we have used the fact that ${\bf X}_{11}^{(0)}={\bf X}_{33}^{(0)}=\frac{a\mathbb{I}}{3d}$ and ${\bf X}_{13}^{(0)}={\bf X}_{31}^{(0)}={\bf X}_{32}^{(0)}=\frac{\mathbb{I}}{3d^3}$. Noticing that ${\bf X}_{12}^{(0)}=\frac{\rho}{3d^2}$ and ${\rm Det}\left({\bf X}_{11}^{(1)}-{\bf X}_{12}^{(1)}\right)={\rm Det}\left(\frac{a\mathbb{I}}{3d}-\frac{\rho}{3d^2}\right)$, we complete %our 
 the proof.
\end{proof}
Therefore, the eigenvalues of the matrix $\rho_{2,1}$ are the union of eigenvalues of matrices, 
$\frac{1}{3}\Big[\frac{\mathbb{I}}{d} -\frac{\rho}{d^2}\Big]$ and $\frac{1}{3}\Big[\frac{\mathbb{I}}{d} + \frac{\rho}{2d^2}\pm  \frac{1}{2}\sqrt{\frac{8\mathbb{I}}{d^6}+\frac{\rho^2}{d^4}}\Big]$. Note that the above method and results directly apply also %applicable 
for the case $\rho_{1,2}$ due to symmetry with respect to the exchange of $M_1$ and $M_2$. %its symmetric nature.

%%%%%%%%%%%%%%%%%%%%%%%
\textbullet{\bf A2. $(3,1)$}: Using the method described in Appendix \ref{diag-bloch-pres}, we find that the characteristic equation for this case takes the form
\begin{align}
{\rm Det}\left(\frac{1}{4}\left[\frac{\mathbb{I}}{d}-\frac{\rho}{d^2}\right]-\lambda \mathbb{I}\right)^{\times 2}\times {\rm Det}\left(\frac{1}{16}\left[\frac{b^2\mathbb{I}}{d^2}+\frac{2b\rho}{d^3}-\frac{3\mathbb{I}}{d^6}\right]\right)=0,
\end{align}
where $b=1-4d\lambda$.
\begin{proof}
  Using Eq.(\ref{det_formula}), the characteristic determinant reduces to 
 ${\rm Det}(\rho_{3,1}-\lambda \mathbb{I})={\rm Det}\left[{\bf X}_{11}^{(3)}\right]{\rm Det}\left[{\bf X}_{22}^{(2)}\right]{\rm Det}\left[{\bf X}_{33}^{(1)}\right]{\rm Det}\left[{\bf X}_{44}^{(0)}\right]$. 
 Again using Eq.(\ref{another-det}), we find ${\bf X}_{11}^{(3)}={\bf X}_{11}^{(2)}-{\bf X}_{12}^{(2)}\left({\bf X}_{22}^{(2)}\right)^{-1}{\bf X}_{21}^{(2)}$, 
 then ${\rm Det}\left[{\bf X}_{11}^{(3)}\right]{\rm Det}\left[{\bf X}_{22}^{(2)}\right]
 ={\rm Det}\left[{\bf X}_{11}^{(2)}{\bf X}_{22}^{(2)}-{\bf X}_{12}^{(2)}{\bf X}_{21}^{(2)}\right]$. 
 Now, we notice that ${\bf X}_{11}^{(2)}={\bf X}_{22}^{(2)}$ and ${\bf X}_{12}^{(2)}={\bf X}_{21}^{(2)}$ which means 
 \begin{align*}
  {\rm Det}\left[{\bf X}_{11}^{(3)}\right]{\rm Det}\left[{\bf X}_{22}^{(2)}\right]={\rm Det}\left[{\bf X}_{11}^{(2)}-{\bf X}_{12}^{(2)}\right]{\rm Det}\left[{\bf X}_{11}^{(2)}+{\bf X}_{12}^{(2)}\right].
 \end{align*}
 At this stage we will consider the product of determinants, ${\rm Det}\left[{\bf X}_{11}^{(2)}+{\bf X}_{12}^{(2)}\right]{\rm Det}\left[{\bf X}_{33}^{(1)}\right]$. Applying Eq.(\ref{det_formula}), we can simplify it as following 
 \begin{align*}
  {\rm Det}\left[{\bf X}_{11}^{(2)}+{\bf X}_{12}^{(2)}\right]{\rm Det}\left[{\bf X}_{33}^{(1)}\right]={\rm Det}\left[\left\{{\bf X}_{11}^{(1)}\right\}^2+{\bf X}_{12}^{(1)}{\bf X}_{11}^{(1)}-2\left\{{\bf X}_{12}^{(1)}\right\}^2\right]\\
  ={\rm Det}\left[{\bf X}_{11}^{(1)}-{\bf X}_{12}^{(1)}\right]{\rm Det}\left[{\bf X}_{11}^{(1)}+2{\bf X}_{12}^{(1)}\right],
 \end{align*}
 where we have used the fact that ${\bf X}_{11}^{(1)}={\bf X}_{33}^{(1)}$ and ${\bf X}_{12}^{(1)}={\bf X}_{13}^{(1)}={\bf X}_{31}^{(1)}={\bf X}_{32}^{(1)}$. Now, the product ${\rm Det}\left[{\bf X}_{11}^{(1)}+2{\bf X}_{12}^{(1)}\right]{\rm Det}\left[{\bf X}_{44}^{(0)}\right]$ simplifies to 
 \begin{align*}
  {\rm Det}\left[{\bf X}_{11}^{(1)}+2{\bf X}_{12}^{(1)}\right]{\rm Det}\left[{\bf X}_{44}^{(0)}\right]={\rm Det}\left[\left\{{\bf X}_{11}^{(0)}\right\}^2+2{\bf X}_{12}^{(0)}{\bf X}_{11}^{(0)}-3\left\{{\bf X}_{14}^{(0)}\right\}^2\right],
 \end{align*}
where we have used the fact that ${\bf X}_{11}^{(0)}={\bf X}_{44}^{(0)}=\frac{b\mathbb{I}}{4d}$ and ${\bf X}_{14}^{(0)}={\bf X}_{41}^{(0)}={\bf X}_{42}^{(0)}=\frac{\mathbb{I}}{4d^3}$. Noticing that ${\bf X}_{12}^{(0)}=\frac{\rho}{4d^2}$ and ${\rm Det}\left[{\bf X}_{11}^{(2)}-{\bf X}_{12}^{(2)}\right]={\rm Det}\left[{\bf X}_{11}^{(1)}-{\bf X}_{12}^{(1)}\right]={\rm Det}\left(\frac{b\mathbb{I}}{4d}-\frac{\rho}{4d^2}\right)$, we complete %our 
the proof.
\end{proof}
Therefore, the eigenvalues of the matrix $\rho_{3,1}$ are the union of eigenvalues of matrices, 
$\frac{1}{4}\Big[\frac{\mathbb{I}}{d} -\frac{\rho}{d^2}\Big]$ with degeneracy $2$ and $\frac{1}{4}\Big[\frac{3\mathbb{I}}{d} + \frac{4\rho}{d^2}\pm \frac{1}{2} \sqrt{\frac{3\mathbb{I}}{d^6}+\frac{\rho^2}{d^4}-\frac{7\mathbb{I}}{16d^2}-\frac{\rho}{2d^3}}\Big]$. Note that due to symmetry the above applies also to the case of $\rho_{1,3}$. %due to its symmetric nature.

%%%%%%%%%%%%%%%%%%%%%%%%%%%%%%
\textbullet{\bf A3. $(3,2)$}: Using the method described in Appendix \ref{diag-bloch-pres}, we find that the characteristic equation for this case takes the form
\begin{align}
{\rm Det}\left(\frac{1}{5}\left[\frac{\mathbb{I}}{d}-\frac{\rho}{d^2}\right]-\lambda \mathbb{I}\right)^{\times 3}\times {\rm Det}\left(\frac{1}{25}\left[\frac{c^2\mathbb{I}}{d^2}+\frac{3c\rho}{d^3}+\frac{2\rho^2}{d^4}-\frac{6\mathbb{I}}{d^6}\right]\right)=0,
\end{align}
where $c=1-5d\lambda$.
\begin{proof}
  Using Eq.(\ref{det_formula}), the characteristic determinant reduces to 
 ${\rm Det}(\rho_{3,2}-\lambda \mathbb{I})=\prod_{k=1}^5 {\rm Det}[{\bf X}_{kk}^{(5-k)}]$.
 %={\rm Det}\left({\bf X}_{11}^{(4)}\right){\rm Det}\left({\bf X}_{22}^{(3)}\right){\rm Det}\left({\bf X}_{33}^{(2)}\right){\rm Det}\left({\bf X}_{44}^{(1)}\right){\rm Det}\left({\bf X}_{55}^{(0)}\right)$. 
 Again using Eq.(\ref{another-det}), we find ${\bf X}_{11}^{(4)}={\bf X}_{11}^{(3)}-{\bf X}_{12}^{(3)}\left({\bf X}_{22}^{(3)}\right)^{-1}{\bf X}_{21}^{(3)}$, 
 then ${\rm Det}\left[{\bf X}_{11}^{(4)}\right]{\rm Det}\left[{\bf X}_{22}^{(3)}\right]
 ={\rm Det}\left[{\bf X}_{11}^{(3)}{\bf X}_{22}^{(3)}-{\bf X}_{12}^{(3)}{\bf X}_{21}^{(3)}\right]$. 
 Now, we notice that ${\bf X}_{11}^{(3)}={\bf X}_{22}^{(3)}$ and ${\bf X}_{12}^{(3)}={\bf X}_{21}^{(3)}$ which means 
 \begin{align*}
  {\rm Det}\left[{\bf X}_{11}^{(4)}\right]{\rm Det}\left[{\bf X}_{22}^{(3)}\right]={\rm Det}\left[{\bf X}_{11}^{(3)}-{\bf X}_{12}^{(3)}\right]{\rm Det}\left[{\bf X}_{11}^{(3)}+{\bf X}_{12}^{(3)}\right].
 \end{align*}
 At this stage we will consider the product of determinants, ${\rm Det}\left[{\bf X}_{11}^{(3)}+{\bf X}_{12}^{(3)}\right]{\rm Det}\left[{\bf X}_{33}^{(2)}\right]$. Applying Eq.(\ref{det_formula}), we can simplify it as follows %ing 
 \begin{align*}
  {\rm Det}\left[{\bf X}_{11}^{(3)}+{\bf X}_{12}^{(3)}\right]{\rm Det}\left[{\bf X}_{33}^{(2)}\right]={\rm Det}\left[\left\{{\bf X}_{11}^{(2)}\right\}^2+{\bf X}_{12}^{(2)}{\bf X}_{11}^{(2)}-2\left\{{\bf X}_{13}^{(2)}\right\}^2\right]\\
  ={\rm Det}\left[{\bf X}_{11}^{(2)}-{\bf X}_{12}^{(1)}\right]{\rm Det}\left[{\bf X}_{11}^{(1)}+2{\bf X}_{12}^{(1)}\right],
 \end{align*}
 where we have used the fact that ${\bf X}_{11}^{(2)}={\bf X}_{33}^{(2)}$ and ${\bf X}_{12}^{(2)}={\bf X}_{13}^{(2)}={\bf X}_{31}^{(2)}={\bf X}_{32}^{(2)}$. Next, the product ${\rm Det}\left[{\bf X}_{11}^{(2)}+2{\bf X}_{12}^{(2)}\right]{\rm Det}\left[{\bf X}_{44}^{(1)}\right]{\rm Det}\left[{\bf X}_{55}^{(0)}\right]$ simplifies to 
 \begin{align*}
  {\rm Det}\left[{\bf X}_{11}^{(2)}+2{\bf X}_{12}^{(2)}\right]{\rm Det}\left[{\bf X}_{44}^{(1)}\right]{\rm Det}\left[{\bf X}_{55}^{(0)}\right]={\rm Det}\left[\left\{{\bf X}_{11}^{(1)}+2{\bf X}_{12}^{(1)}\right\}{\bf X}_{44}^{(1)}-3\left\{{\bf X}_{14}^{(1)}\right\}^2\right]{\rm Det}\left[{\bf X}_{55}^{(0)}\right]\\
  ={\rm Det}\left[\left\{{\bf X}_{11}^{(1)}+2{\bf X}_{12}^{(1)}\right\}\left\{{\bf X}_{44}^{(0)}{\bf X}_{55}^{(0)}-{\bf X}_{45}^{(0)}{\bf X}_{54}^{(0)}\right\}-3{\bf X}_{14}^{(1)}\left\{{\bf X}_{14}^{(0)}{\bf X}_{55}^{(0)}-{\bf X}_{15}^{(0)}{\bf X}_{54}^{(0)}\right\}\right],
 \end{align*}
 where in the first line we have used the fact that ${\bf X}_{14}^{(1)}={\bf X}_{41}^{(1)}={\bf X}_{42}^{(1)}$. 
We know that ${\bf X}_{44}^{(0)}=\frac{c\mathbb{I}}{5d}$, ${\bf X}_{45}^{(0)}={\bf X}_{54}^{(0)}=\frac{\rho}{5d^2}$ and ${\bf X}_{14}^{(0)}={\bf X}_{15}^{(0)}=\frac{\mathbb{I}}{5d^3}$. 
After few tedious algebraic steps, we find that ${\bf X}_{11}^{(1)}=\frac{c\mathbb{I}}{5d}-\frac{\mathbb{I}}{5cd^3}$, ${\bf X}_{12}^{(1)}=\frac{\rho}{5d^2}-\frac{\mathbb{I}}{5cd^3}$ and ${\bf X}_{14}^{(1)}=\frac{\mathbb{I}}{5d^3}-\frac{\rho}{5cd^4}$. Putting all these in the above equation, we find the following simplification
\begin{align*}
 {\rm Det}\left[{\bf X}_{11}^{(2)}+2{\bf X}_{12}^{(2)}\right]{\rm Det}\left[{\bf X}_{44}^{(1)}\right]{\rm Det}\left[{\bf X}_{55}^{(0)}\right]={\rm Det}\left(\frac{1}{5}\left[\frac{\mathbb{I}}{d}-\frac{\rho}{d^2}\right]-\lambda \mathbb{I}\right){\rm Det}\left(\frac{1}{25}\left[\frac{c^2\mathbb{I}}{d^2}+\frac{3c\rho}{d^3}+\frac{2\rho^2}{d^4}-\frac{6\mathbb{I}}{d^6}\right]\right).
\end{align*}
Noticing that ${\rm Det}\left({\bf X}_{11}^{(2)}-{\bf X}_{12}^{(2)}\right)={\rm Det}\left({\bf X}_{11}^{(3)}-{\bf X}_{12}^{(3)}\right)={\rm Det}\left(\frac{c\mathbb{I}}{5d}-\frac{\rho}{5d^2}\right)$, we complete our proof.
\end{proof}
Therefore, the eigenvalues of the matrix $\rho_{3,2}$ are the union of eigenvalues of matrices, 
$\frac{1}{5}\Big[\frac{\mathbb{I}}{d} -\frac{\rho}{d^2}\Big]$ with degeneracy $3$ and $\frac{1}{5}\Big[\frac{\mathbb{I}}{d} + \frac{3\rho}{2d^2}\pm \frac{1}{2} \sqrt{\frac{24\mathbb{I}}{d^6}+\frac{(9-6d-8\rho)\rho}{d^4}}\Big]$. 
%Note that the above method is also applicable for the case $\rho_{2,3}$ due to its symmetric nature.
Due to symmetry the above results apply also for the case of $\rho_{2,3}$.

The reduced density matrix for the control system after the evolution can be calculated using the prescription given in Ref.\cite{PhysRevLett.120.120502} and is given by
\begin{align}
 \tilde{\rho_{c}}(M)=\frac{1}{M}\left\{\mathbb{I}_c+\frac{1}{d^2}\sum_{i\neq j}^{M-1}\ketbra{i}{j}
 %+\frac{1}{d^3}\sum_{\ell,\mu}\left(\vert \ell \rangle\langle \mu \vert_c +\vert \mu \rangle\langle \ell \vert_c\right)
 \right\},
\end{align}
whose eigenvalues are $\frac{1}{M}\left(1-\frac{1}{d^2}\right)$ with degeneracy $M-1$ and $\frac{1}{M}\left(1+\frac{M-1}{d^2}\right)$ with degeneracy one.

Note that we can retrieve the special cases of the cyclic orders by replacing either $M_1=0$ or $M_2=0$. 

\subsection{N=4 case}\label{4chanState}
%For four channels there are six ($(4-1)!=6$) cosets of cyclic causal orders. Let us denote $M_{\eta}$ as the number of causal orders in a coset where \tb{ $M_{\eta}\in[1,4]$ }for each $\eta\in[1,6]$. Let us consider that the target state is $\rho$. Now, if we consider $M = (\sum_{\eta}M_{\eta})$ causal orders in a quantum SWITCH, the output state will have -- a) \tb{within each cyclic coset, the diagonal terms are }proportional to $\frac{\mathbb{I}}{d}$ and off-diagonal terms are proportional to $\frac{\rho}{d^2}$; and  b) cross-coset off-diagonal terms are proportional to $\frac{\mathbb{I}}{d^3}$ as well as $\frac{\rho}{d^4}$. 
%
To show that $\frac{\rho}{d^4}$ terms occurs in off-diagonal, we will pick two instances:
\begin{align*}
 \frac{1}{d^8}\sum_{ijk\ell} U_jU_kU_{\ell} U_i\left(\rho U_j^\dagger U_{\ell}^\dagger \right) U_i^\dagger U_k^\dagger
   =\frac{1}{d^8}\cdot d^2\sum_{jk\ell}U_jU_kU_{\ell}\left({\rm Tr}[\rho U_j^\dagger U_{\ell}^\dagger] \frac{\mathbb{I}}{d}\right) U_k^\dagger \\
   = \frac{1}{d^6}\sum_{jk}U_jU_k\left(\rho U_j^\dagger  \right) U_k^\dagger
   =\frac{1}{d^6}\cdot d^2\sum_{j} U_j{\rm Tr}[\rho U_j^\dagger]\frac{\mathbb{I}}{d} =\frac{\mathbb{\rho}}{d^4},
\end{align*}
where we have used the fact that $\frac{1}{d}\sum_{k}\left({\rm Tr}[\rho U_k^\dagger] \right) U_k=\rho$. And other off-diagonal term,
\begin{align*}
\frac{1}{d^8}\sum_{ijk\ell} U_j U_{\ell}U_k U_i\left(\rho U_j^\dagger  U_k^\dagger U_{\ell} \right) U_i^\dagger 
   =\frac{1}{d^8}\cdot d^2\sum_{jk\ell}U_j U_{\ell}U_k\left({\rm Tr}[\rho U_j^\dagger U_{\ell}^\dagger U_k^\dagger] \frac{\mathbb{I}}{d}\right)  \\
   = \frac{1}{d^6}\sum_{j\ell}U_j U_{\ell}\left(\rho U_j^\dagger  \right) U_{\ell}^\dagger
   =\frac{1}{d^6}\cdot d^2\sum_{j} U_j{\rm Tr}[\rho U_j^\dagger]\frac{\mathbb{I}}{d} =\frac{\mathbb{\rho}}{d^4},
\end{align*}
where we have used the fact that $\frac{1}{d}\sum_{k}\left({\rm Tr}[\rho U_j^\dagger U_{\ell}^\dagger U_k^\dagger] \right) U_k=\rho U_j^\dagger U_{\ell}^\dagger$.

%%%%%%%%%%%%%%%%%

%We consider a specific way of selecting causal orders for each cosets, eg., the first entry of each cosets are linked with the zeroth

%\tb{ We use a particular way of listing causal orders for each coset. Namely, the first entry of each coset is linked with %the zeroth
%the order $\Lambda_1\Lambda_2\Lambda_3\Lambda_4$ by some %non-cyclic 
%permutation of the last three labels.} In this way, the first element in each coset is starting with $\Lambda_1$. And rest is %will be
%constructed using cyclic permutations from first element of corresponding coset. Using this setup, we find that the output state after evolution is given by
Using the ordering within cosets described in the main text, we find that the output state after evolution is given by
\begin{align}
 \rho_{M_{\eta}}=\frac{1}{M}\left\{\mathbb{I}_c\otimes\frac{\mathbb{I}}{d}+ L_{M_{\eta}}\otimes \frac{\rho}{d^2}+ B_{M_{\eta}}\otimes \frac{\mathbb{I}}{d^3} +Q_{M_{\eta}}\otimes \frac{\rho}{d^4}   \right\},
 \label{4chanellsA}
\end{align}
where the matrices, $L$, $B$ and $Q$ are given by
\begin{align*}
{\bf L}=\left(
  \begin{array}{ccccc}
    S_{M_1\times M_1} & {\bf 0}_{M_1\times M_2} & {\bf 0}_{M_1\times M_3} & \cdots & {\bf 0}_{M_1\times M_6}\\
    {\bf 0}_{M_2\times M_1} & S_{M_2\times M_2} & {\bf 0}_{M_2\times M_3} & \cdots & {\bf 0}_{M_2\times M_6}\\
    \vdots & \vdots & \ddots & \vdots & \vdots  \\
     {\bf 0}_{M_5\times M_1} & {\bf 0}_{M_5\times M_2} & \cdots & S_{M_5\times M_5} & {\bf 0}_{M_5\times M_6}\\
     {\bf 0}_{M_6\times M_1} & {\bf 0}_{M_6\times M_2} & \cdots &  {\bf 0}_{M_6\times M_5} & S_{M_6\times M_6} \\
  \end{array}
\right),
\end{align*}
and for sake of simplicity we are writing the form of $B$ and $Q$ when $M=4!$, i.e.,
\begin{align*}
{\bf B}=\left(
  \begin{array}{cccccc}
    {\bf 0} & B_1 & B_2 & B_3 & B_4 & B_7\\
    B_1 & {\bf 0} & B_3 & B_6 & B_2 & B_4\\
    B_2 & B_3 & {\bf 0} & B_4 & B_5 & B_1  \\
    B_3 & B_6 & B_4 & {\bf 0} & B_1 & B_2\\
    B_4 & B_2 & B_5 &  B_1 & {\bf 0} & B_3\\
    B_7 & B_4 & B_1 &  B_2 & B_3 & {\bf 0}\\
  \end{array}
\right),\hspace{0.25cm}
{\bf Q}=\left(
  \begin{array}{cccccc}
    {\bf 0} & Q_1 & Q_2 & Q_3 & Q_4 & Q_7\\
    Q_1 & {\bf 0} & Q_3 & Q_6 & Q_2 & Q_4\\
    Q_2 & Q_3 & {\bf 0} & Q_4 & Q_5 & Q_1  \\
    Q_3 & Q_6 & Q_4 & {\bf 0} & Q_1 & Q_2\\
    Q_4 & Q_2 & Q_5 &  Q_1 & {\bf 0} & Q_3\\
    Q_7 & Q_4 & Q_1 &  Q_2 & Q_3 & {\bf 0}\\
  \end{array}
\right),\hspace{0.25cm}
\end{align*}
where $B_i$ and $Q_i$ are defined below, 
\begin{align*}
 B_1=\left(\begin{array}{cccc}
    1 & 0 & 1 & 1 \\
    0 & 1 & 0 & 0\\
    1 & 0 & 1 & 1   \\
     1 & 0 & 1 & 1 \\
  \end{array}
\right),\hspace{0.25cm}
B_5=\left(\begin{array}{cccc}
    1 & 1 & 0 & 0 \\
    1 & 1 & 0 & 0\\
    0 & 0 & 1 & 1   \\
     0 & 0 & 1 & 1 \\
  \end{array}
\right);
\hspace{0.25cm}
 Q_1=\left(\begin{array}{cccc}
    0 & 1 & 0 & 0 \\
    1 & 0 & 1 & 1\\
    0 & 1 & 0 & 0   \\
     0 & 1 & 0 & 0 \\
  \end{array}
\right),
\hspace{0.25cm}
Q_5=\left(\begin{array}{cccc}
    0 & 0 & 1 & 1 \\
    0 & 0 & 1 & 1\\
    1 & 1 & 0 & 0   \\
     1 & 1 & 0 & 0 \\
  \end{array}
\right),
\end{align*}
and $B_p=\pi_p [B_1]$ for $p=2,3,4$ and $B_p=\pi_p [B_5]$ for $p=6,7$; and similarly, $Q_p=\pi_p [Q_1]$ for $p=2,3,4$ and $Q_p=\pi_p [Q_5]$ for $p=6,7$, where $\pi_p$ are permutation operators which permutes among the columns as well as rows of the target matrix. For our scenario, $\pi_2=\pi^{row}_{1\leftrightarrow 2}\cdot\pi^{col}_{1\leftrightarrow 2}$, $\pi_3=\pi_7=\pi^{row}_{2\leftrightarrow 4}\cdot\pi^{col}_{2\leftrightarrow 4}$, and $\pi_4=\pi_6=\pi^{row}_{2\leftrightarrow 3}\cdot\pi^{col}_{2\leftrightarrow 3}$, where $i\leftrightarrow j$ denotes exchange between the specific labels. Therefore, we find that in each row (and column) in $\rho_{M_{\eta}}$, there are $12$ entries proportional to $\frac{\mathbb{I}}{d^3}$ and $8$ entries proportional to $\frac{\rho}{d^4}$. Note however that our analysis of Holevo quantity is invariant under arbitrary arrangement of elements in a particular coset. The reduced density matrix for the control qubit after the evolution can be calculated using the prescription given in Ref.\cite{PhysRevLett.120.120502} and is given by
\begin{align}
 \tilde{\rho_{c}}(M_{\eta})=\frac{1}{M}\left\{\mathbb{I}_c+\frac{1}{d^2}(L_{M_{\eta}}+B_{M_{\eta}})+\frac{1}{d^4} Q_{M_{\eta}}
 %+\frac{1}{d^3}\sum_{\ell,\mu}\left(\vert \ell \rangle\langle \mu \vert_c +\vert \mu \rangle\langle \ell \vert_c\right)
 \right\}.
\end{align}
% whose eigenvalues are $\frac{1}{M}\left(1-\frac{1}{d^2}\right)$ with degenracy $M-1$ and $\frac{1}{M}\left(1+\frac{M-1}{d^2}\right)$ with degeneracy one.

%It is hard (not impossible!) to 
It would be too tedious to diagonalize analytically the output matrix in Eq.(\ref{4chanellsA}) for $M=4!$ for arbitrary $d$. Instead, we will %resort to 
use numerical methods to diagonalize the output state in order to calculate the Holevo quantity of effective channel. However, there are specific cases, where we can easily take the analytical routes. Specifically, we are considering some non-trivial situations, like, 

{\bf A1. $M=8$}: We are considering a scenario where most of the off-diagonal elements are proportional to $\rho$. Such a situation occurs when $M_3=M_5=4$ and other $M_{\eta}=0$. Therefore, the output density matrix is given by
\begin{align}
 \rho_{4,4,\vec{0}}=\frac{1}{8}\left(\mathbb{I}_c\otimes\frac{\mathbb{I}}{d}+ L_{4,4}\otimes \frac{\rho}{d^2}+ B_{4,4}\otimes \frac{\mathbb{I}}{d^3} +Q_{4,4}\otimes \frac{\rho}{d^4}   \right),
\end{align}
where the coefficient matrices $(L$, $B$, $Q)$ are 
\begin{align*}
  L_{4,4}=\left(
  \begin{array}{cc}
    S_{4\times 4} & {\bf 0}_{4\times 4} \\
    {\bf 0}_{4\times 4} & S_{4\times 4} \\
  \end{array}
\right),
 B_{4,4}=\left(
  \begin{array}{cc}
    {\bf 0}_{4\times 4} & B_5 \\
    B_5  & {\bf 0}_{4\times 4} \\
  \end{array}
\right),\hspace{0.25cm} \mbox{and}\hspace{0.25cm}
Q_{4,4}=\left(
  \begin{array}{cc}
    {\bf 0}_{4\times 4} & Q_5 \\
    Q_5 & {\bf 0}_{4\times 4} \\
  \end{array}
\right).
\end{align*}
As $[L_{4,4},B_{4,4}]=[L_{4,4},Q_{4,4}]=[B_{4,4},Q_{4,4}]=0$, the matrices $L$, $B$ and $Q$ are simultaneously diagonalizable, i.e., there exists a unitary matrix $U_{4,4}$ such that $U^\dagger L U={\rm Diag}(3,3, -1,\cdots, -1)$ and $U^\dagger B U=U^\dagger Q U={\rm Diag}(-2,2, 0,\cdots, 0)$.

{\bf A2. $M=6$}: We can choose many scenarios here with different $M_{\eta}$. The most trivial case can occur when  $M_{\eta}=1$ $\forall \eta$. However, there exists two interesting scenarios: %, one such example for each are --
1) $(M_1=3,M_4=3)$ and 2) $(M_3=4,M_5=2)$.

{\em Example 1}: We consider those causal order for $(M_1=3,M_4=3)$ such that the output state is exactly that of the $(N=3, M=6)$ given in Eq.(\ref{N3M6}). And we know that the output state is exactly diagonolizable using simultaneously diagonalisation method.

{\em Example 2}: Here we choose those causal orders for $(M_3=4,M_5=2)$ such that we get the following output state 
\begin{align*}
 \rho_{4,2,\vec{0}}=\frac{1}{6}\left(\mathbb{I}_c\otimes\frac{\mathbb{I}}{d}+ L_{4,2}+\otimes \frac{\rho}{d^2}+ B_{4,2}\otimes \frac{\mathbb{I}}{d^3} +Q_{4,2}\otimes \frac{\rho}{d^4}   \right),
\end{align*}
where the coefficient matrices $(L$, $B$, $Q)$ are 
\begin{align*}
  L_{4,4}=\left(
  \begin{array}{cc}
    S_{4\times 4} & {\bf 0}_{4\times 2} \\
    {\bf 0}_{2\times 4} & S_{2\times 2} \\
  \end{array}
\right),
 B_{4,4}=\left(
  \begin{array}{cc}
    {\bf 0}_{4\times 4} & \tilde{B}_5^T \\
    \tilde{B}_5  & {\bf 0}_{2\times 2} \\
  \end{array}
\right),\hspace{0.25cm} \mbox{and}\hspace{0.25cm}
Q_{4,4}=\left(
  \begin{array}{cc}
    {\bf 0}_{4\times 4} & \tilde{Q}_5^T \\
    \tilde{Q}_5 & {\bf 0}_{2\times 2} \\
  \end{array}
\right),
\end{align*}
with $\tilde{B}_5=\big(\begin{smallmatrix}
  0 & 0 & 1 & 1\\
  0 & 0 & 1 & 1
\end{smallmatrix}\big)$ 
and 
$\tilde{Q}_5=\big(\begin{smallmatrix}
  1 & 1 & 0 & 0 \\
  1 & 1 & 0 & 0 
\end{smallmatrix}\big)$. The output state in this case can be diagonalized using the method given in Appendix \ref{diag-bloch-pres}.

{\em Example.3}: The trivial scenario can occur when $M_{\eta}=1$ $\forall \eta$ where all the off-diagonal terms are proportional to $\mathbb{I}$. However, we consider another scenario where not all off-diagonal elements are proportional to $\mathbb{I}$ and the output state is given by
\begin{align}
\rho_{\vec{1}}=\frac{1}{6}\left(
  \begin{array}{cccccc}
    \frac{\mathbb{I}}{d} & \frac{\rho}{d^4}  & \frac{\rho}{d^4}  & \frac{\rho}{d^4}  & \frac{\rho}{d^4}  & \frac{\rho}{d^4} \\
   \frac{\rho}{d^4}  & \frac{\mathbb{I}}{d} & \frac{\mathbb{I}}{d^3} & \frac{\rho}{d^4}  & \frac{\mathbb{I}}{d^3} & \frac{\rho}{d^4} \\
   \frac{\rho}{d^4}  & \frac{\mathbb{I}}{d^3} & \frac{\mathbb{I}}{d} & \frac{\mathbb{I}}{d^3} & \frac{\rho}{d^4}  & \frac{\mathbb{I}}{d^3}  \\
    \frac{\rho}{d^4}  & \frac{\rho}{d^4}  & \frac{\mathbb{I}}{d^3} &\frac{\mathbb{I}}{d} & \frac{\mathbb{I}}{d^3} & \frac{\rho}{d^4} \\
    \frac{\rho}{d^4}  & \frac{\mathbb{I}}{d^3} & \frac{\rho}{d^4}  &  \frac{\mathbb{I}}{d^3} & \frac{\mathbb{I}}{d} & \frac{\mathbb{I}}{d^3}\\
    \frac{\rho}{d^4}  & \frac{\rho}{d^4}  & \frac{\mathbb{I}}{d^3} &  \frac{\rho}{d^4}  & \frac{\mathbb{I}}{d^3} & \frac{\mathbb{I}}{d}\\
  \end{array}
\right).
\label{6app_didnot}
\end{align}
This matrix can be diagonalized using the method presented in Appendix \ref{diag-bloch-pres}.

\section{Determinant of block matrices}\label{diag-bloch-pres}
In order to find eigenvalues of the $dM\times dM$ block matrix, $\rho_{out}$, in the main text we need to find how its determinant factorizes into determinant of small matrices as 
discussed in the Ref.\cite{10.2307/3620776,powell2011calculating}. We will state the Lemma from the Ref.\cite{10.2307/3620776,powell2011calculating} below,\\

\begin{lemma}
Let ${\bf A}$ be an $pN\times pN$ complex matrix partitioned into $N^2$-Blocks, each of size $p\times p$, i.e., 
\begin{equation*}
{\bf A}=\left(
  \begin{array}{cccc}
    A_{11} & A_{12}  & \cdots & A_{1N} \\
    A_{21}  & A_{22} &  \cdots & A_{2N} \\
    \vdots  & \vdots & \ddots & \vdots \\
    A_{N1} & A_{N2} & \cdots & A_{NN} \\
  \end{array}
\right)
_{pN\times pN},
\end{equation*}
then its determinant is given by 
\begin{align}
{\rm Det}[ {\bf A}]=\prod_{k=1}^N {\rm Det}[{\bf X}_{kk}^{(N-k)}],
\label{det_formula}
\end{align}
where ${\bf X}^{(i)}$ are defined as
\begin{align*}
&{\bf X}_{ij}^{(0)}=A_{ij},\\
&{\bf X}_{ij}^{(k)}=A_{ij}- \vec{b}_{i,N-k+1}^T\tilde{{\bf A}}_k^{-1}\vec{a}_{N-k+1,j}, \hspace{1cm} k\geq 1,
\end{align*}
with $\vec{a}_{ij}=(A_{ij}, A_{i+1,j},\cdots,A_{Nj})^T$, $\vec{b}^T=(A_{ij},A_{i,j+1},\cdots,A_{iN})$, and $\tilde{{\bf A}}_k$ being the $k\times k$ block matrix formed from the lower right corner of ${\bf A}$. The author in Ref.\cite{powell2011calculating} also notices that 
\begin{align}
{\bf X}_{ij}^{(k+1)}={\bf X}_{ij}^{(k)}- {\bf X}_{i,N-k}^{(k)}\Big({\bf X}_{N-k,N-k}^{(k)}\Big)^{-1}{\bf X}_{N-k,j}^{(k)}.
\label{another-det}
\end{align}
\end{lemma}
Equipped with the above Lemma, we will try to find the eigenvalues of matrix $\rho_{M}$.

\end{widetext}

%\bibliography{Nchannel_ref}
\end{document}